\UseRawInputEncoding
\documentclass[aip,amsmath,amssymb,reprint]{revtex4-1}

\usepackage{graphicx}% Include figure files
\usepackage{dcolumn}% Align table columns on decimal point
\usepackage{bm}% bold math
\usepackage[english]{babel}

\usepackage{mathptmx}
\usepackage{etoolbox}

\makeatletter
\def\@email#1#2{%
 \endgroup
 \patchcmd{\titleblock@produce}
  {\frontmatter@RRAPformat}
  {\frontmatter@RRAPformat{\produce@RRAP{*#1\href{mailto:#2}{#2}}}\frontmatter@RRAPformat}
  {}{}
}%
\makeatother

\usepackage{epsfig}
\usepackage{color}

\usepackage{stackengine}
\usepackage{multirow}

\renewcommand{\(}{\left(} 
\renewcommand{\)}{\right)} 
\renewcommand{\[}{\left[} 
\renewcommand{\]}{\right]}

\newcommand{\nn}{\nonumber} 
\newcommand{\intl}{\int\limits}

\newcommand{\etac}{\eta_\mathrm{c}}
\newcommand{\Dc}{D_\mathrm{c}}
\newcommand{\Cc}{C_\mathrm{c}}

\newcommand{\bs}[1]{\boldsymbol{#1}}
\newcommand{\lr}[1]{\langle{#1}\rangle}

\begin{document}

\title{A self-consistent analytical theory for rotator networks under stochastic forcing: effects of intrinsic noise and common input}

\author{Jonas Ranft}
\email{jonas.ranft@ens.psl.eu}
\affiliation{Institut de Biologie de l'ENS, Ecole Normale Sup{\'e}rieure, CNRS, Inserm, Universit{\'e} PSL, 46 rue d'Ulm, 75005 Paris, France}
\author{Benjamin Lindner}
\affiliation{Bernstein Center for Computational Neuroscience Berlin, Philippstra{\ss}e 13, Haus 2, 10115 Berlin, Germany and Physics Department of Humboldt University Berlin, Newtonstra{\ss}e 15, 12489 Berlin, Germany}

\date{\today}

\begin{abstract}
Despite the incredible complexity of our brains' neural networks, theoretical descriptions of neural dynamics have led to profound insights into possible network states and dynamics. It remains challenging to develop theories that apply to spiking networks and thus allow one to characterize the dynamic properties of biologically more realistic networks. Here, we build on recent work by van Meegen \& Lindner who have shown that ``rotator networks,'' while considerably simpler than real spiking networks and therefore more amenable to mathematical analysis, still allow to capture dynamical properties of networks of spiking neurons. This framework can be easily extended to the case where individual units receive uncorrelated stochastic input which can be interpreted as intrinsic noise. However, the assumptions of the theory do not apply anymore when the input received by the single rotators is strongly correlated among units. As we show, in this case the network fluctuations become significantly non-Gaussian, which calls for a reworking of the theory. Using a cumulant expansion, we develop a self-consistent analytical theory that accounts for the observed non-Gaussian statistics. Our theory provides a starting point for further studies of more general network setups and information transmission properties of these networks. 
\end{abstract}

\maketitle

\begin{quotation}

Not surprisingly, networks of recurrently connected neurons
, such as e.g.~found in the human cortex, 
are capable to support a very rich repertoire of dynamical states. A remarkable feature of such networks is that they can exhibit asynchronous, irregular activity even in the case of 
%completely 
deterministic dynamics, where the disorder due to the random connectivity between neurons 
%effectively 
gives rise to temporal fluctuations of their activity. Understanding the statistics and dynamical properties of network states by mathematical analysis 
of models of recurrent networks 
remains a challenging problem especially for networks of spiking neurons, for which analytical expressions e.g.~for the network noise autocorrelation function are hard to obtain. For these reasons, we investigate here a much simpler model class of recurrently connected oscillators that nevertheless has been shown to give a reasonable account of the statistics of spiking networks, and that allows one to obtain in comparison much simpler self-consistent equations for the network statistics. Using this model, we study the effects of intrinsic noise uncorrelated among oscillators and 
%common noise shared among oscillators. 
shared common noise.
%Whereas intrinsic noise leads to a relatively straightforward extension of the existing theory for autonomous networks without noise, the effect of common noise is more difficult to capture as the network fluctuations become non-Gaussian even if both  the self-generated network noise in the absence of input and the additional common noise are Gaussian. 
Interestingly, we find that in the presence of common noise the network fluctuations become non-Gaussian even if both the self-generated network noise in the absence of noise and the additional common noise are Gaussian. To account for the non-Gaussian statistics, we develop a self-consistent theory that includes higher cumulants of the network fluctuations. Our theory provides a satisfactory description of the observed power spectra and autocorrelation functions of the network fluctuations and rotator dynamics, and can serve as a starting point for further investigations of networks in which non-Gaussian statistics due to input correlations might play a role.

\end{quotation}

\section{Introduction}

Networks of recurrently coupled oscillators have been studied in a wide variety of contexts, representing spatially inhomogeneous chemical reactions~\cite{Kuramoto:1984wo}, neural networks~\cite{Ermentrout:1998ct,Grytskyy:2013fr,diVolo:2018ep}, 
%% JR: not Grytskyy et al. do not study oscillators actually...
%% ... maybe di Volo & Torcini. PRL (2018) ?
%% ... not really satisfying ...
or power grids~\cite{Filatrella:2008co}. A classic paradigm has been and continues to be the Kuramoto model, which allowed many analytical insights in e.g.~the synchronization dynamics depending on the disorder in the system~\cite{Kuramoto:1984wo,Pikovsky:2001vm,Acebron:2005ik}. More generally, networks of recurrently connected units can be in a regime where they exhibit apparently noisy dynamics despite being entirely deterministic, a feature which has received a lot of attention in the context of neural networks~\cite{Sompolinsky:1988cs,Brunel:2000th,Lerchner:2006ko,Kriener:2008ca,Dummer:2014cz,Ostojic:2014kd,vanMeegen:2018hj,Pena:2018gl,Vellmer:2019fl}. While such noisy dynamics can easily be observed  e.g.~in spiking network models where it manifests itself in the ``asynchronous state'' \cite{Renart:2010hj}, it is considerably harder to analytically determine the effective noise characteristics for such models. In order to thoroughly characterize the internally-generated noise as a function of network parameters and understand its impact e.g.~on the statistics of individual units in the network, it seems important to have models at one's disposal that capture the phenomenon while still allowing one to gain analytical insights.  

Recently, van Meegen \& Lindner~\cite{vanMeegen:2018hj} introduced the so-called ``rotator network'' of recurrently coupled rotators, in which the coupling between the oscillating units is an arbitrary function of the ``presynaptic'' oscillator phases. Despite being a strongly simplified description of real spiking networks, the authors showed that the model still allows to capture dynamical properties of model networks of spiking neurons. Importantly, for the rotator network one can self-consistently calculate the statistics of the internally-generated fluctuations (network noise). Here, we will build upon this work to investigate the influence of additional intrinsic noise sources at the level of the individual units as well as shared common input. The additional input leads to modifications of the self-consistent equation for the autocorrelation function of the network noise and impacts the individual units' statistics. When the external inputs are uncorrelated across the units, the required modifications of the theory are straightforward. However, if there are substantial correlations in the input across units, basic assumptions of the existing theory break down and require a more substantial revision with regard to the non-Gaussian statistics of the network noise. Extending the theory accordingly, we aim to go a step further towards increased biological realism in the context of neural networks while maintaining the analytical tractability of the model. 
%More generally to develop a deeper understanding of the dynamics of networks recurrently coupled oscillators. 

After briefly introducing the basic model and reviewing previous results in section~\ref{sec:previous}, we extend the rotator network model to the case where individual units are subject to additional intrinsic noise in section~\ref{sec:intrinsic}. Intrinsic noise can e.g.~be caused by stochastic ion channel openings in the case of neural networks, or by fluctuating inputs or loads in power grids. In section~\ref{sec:common}, we discuss the effect of shared external input on all units. Whereas the internally-generated network noise remains Gaussian in the presence of intrinsic noise only, it becomes non-Gaussian in the presence of correlated inputs. This is captured by a modified self-consistent equation that takes into account higher cumulants of the internally-generated noise. We conclude in section~\ref{sec:discussion} with a summary of our results and a brief discussion of their implications.

\section{Previous results: network dynamics without external inputs}
\label{sec:previous}

In a previous study, van Meegen \& Lindner introduced a simple model of a recurrent rotator network, where individual units are described as phase variables $\theta_m$, each driven by an intrinsic oscillation frequency $\omega_m$ and receiving input from the other units in the network,
\begin{equation}
\label{eq:vML-model}
\dot \theta_m = \omega_m + \sum_n K_{mn} f(\theta_n) ,
\end{equation}
where the recurrent input is specified by the matrix of connection weights $K_{mn}$ between units $n$ and $m$, and the coupling function $f(\cdot)$ that depends only on the `input phases' $\theta_n$. To be specific, we will assume $K_{mn}$ to be a Gaussian random matrix with $\langle K_{mn} \rangle=0$ and $\langle K_{mn} K_{m'n'} \rangle=\frac{K^2}{N}\delta_{mm'}\delta_{nn'}$ unless specified otherwise, but see section~\ref{sec:discussion} for a discussion of other choices for the connectivity. Based on dynamical mean-field theory, Eq.~\eqref{eq:vML-model} can be interpreted as a stochastic dynamics for the $\theta_m$ with an internally generated network noise $\xi_m$,
\begin{equation}
\label{eq:xi_m}
\xi_m = \sum_n K_{mn} f(\theta_n),
\end{equation}
the statistics of which has to be determined self-consistently. The relative simplicity of the model allows one to derive an analytical solution for the 
noise autocorrelation function $C_{\xi}(\tau)\equiv\langle \xi_m(t)\xi_m(t+\tau)\rangle$. 
To this end, following van Meegen \& Lindner, we introduce the auxiliary function
%rotator autocorrelation function $C_{x_m}(\tau)\equiv\langle x_m^*(t)x_m(t+\tau)\rangle = \exp[i\omega_m\tau - \Lambda(\tau)]$, 
\begin{equation}
\label{eq:Lambda}
\Lambda(\tau) = \int_0^\tau dt(\tau-t)C_\xi(t) ,
\end{equation}
which implies
\begin{equation}
\label{eq:Cxi_from_Lambda_ddot}
C_\xi(\tau) = \ddot \Lambda(\tau) .
\end{equation}
In the case of the autonomous network described by Eq.~\eqref{eq:vML-model}, the function $\Lambda(\tau)$ is a solution of the ordinary differential equation 
\begin{equation}
\label{eq:vML-Lambda}
\ddot\Lambda(\tau) = K^2\sum_l |A_l|^2 \Phi(l\tau)\exp[-l^2\Lambda(\tau)].
\end{equation}
Here, $A_l$ are the coefficients of a Fourier expansion of the coupling function, $f(\theta) = \sum_l A_l \exp[il\theta]$, and $\Phi(x)=\langle \exp[i\omega_nx]\rangle_{\omega_n}$ is the characteristic function of the intrinsic frequencies. The above equation for $\Lambda$ follows from a self-consistency condition for the noise autocorrelation, where the average is taken over realizations of the connectivity matrix $K_{mn}$, intrinsic frequencies $\omega_n$, initial conditions, and the effective noise $\xi_m$ in a self-consistent manner.

We would like to emphasize that, while the relations of $\Lambda(\tau)$ in Eqs.~\eqref{eq:Lambda} and~\eqref{eq:Cxi_from_Lambda_ddot} hold true throughout this manuscript, the dynamical equation that one obtains for $\Lambda$ will depend on the situation considered (absence or presence of intrinsic and/or common noise) and, for the most involved case of common noise, on the level of approximation used. It would be possible to add a label to the function $\Lambda$ that would distinguish these different cases, but we abstain from doing so for the ease of notation.

We quickly recapitulate the derivation  of van Meegen \& Lindner for reference, also because our new results will be obtained along similar lines. Using the Gaussian statistics when averaging over the connectivity matrix, one first obtains 
\begin{subequations}
\label{eq:vML_Cxi_deriv1}
\begin{align}
C_{\xi}(\tau) &= \sum_{n,n'} \langle K_{mn}K_{mn'} f(\theta_n(t))f(\theta_{n'}(t+\tau))\rangle_{K,\bs{\omega},\bs{\theta_0}} \\
&= \sum_{n,n'} \langle K_{mn}K_{mn'} \rangle_K \langle f(\theta_n(t))f(\theta_{n'}(t+\tau))\rangle_{K,\bs{\omega},\bs{\theta_0}} \\
&= \frac{K^2}{N} \sum_n \langle f(\theta_n(t))f(\theta_n(t+\tau))\rangle_{K,\bs{\omega},\bs{\theta_0}} . %\\
%&= K^2 \langle f(\theta_n(t))f(\theta_n(t+\tau))\rangle_{K,\bs{\omega},\bs{\theta_0}},
\end{align}
\end{subequations}
%where the last equation holds because the terms of the sum over $n$ do not explicitly depend on the index after averaging. 
Note that the disorder of the connectivity matrix in principle still contributes to the fluctuating dynamics of the $\theta_n(t)$. The self-consistency ansatz for statistics of the internally-generated network fluctuations now consists in considering the effective stochastic dynamics $\dot\theta_m=\omega_m+\xi_m$ for the rotators in the network and to average over realizations of the network noise $\xi_m(t)$ in the following. Using the Fourier expansion of $f(\cdot)$ and $\theta_m(t) = \theta_m(t_0) + \int_{t_0}^t dt'\dot\theta_m(t')$, one thus obtains
%\begin{subequations}
%\label{eq:vML_Cxi_deriv2}
%\begin{align}
%\frac{C_{\xi}(\tau)}{K^2} &=  \sum_{l,l'} A_l A_{l'} \langle  e^{i l \theta_n(t) + il'\theta_n(t+\tau)}\rangle_{\bs{\xi},\bs{\omega},\bs{\theta_0}} \\
%&=  \sum_{l,l'} A_l A_{l'} \langle e^{i(l+l')\theta_n(t_0)} \rangle_{\bs{\theta_0}} \times \nonumber \\
%&\qquad \qquad \langle e^{i l \int_{t_0}^t dt'\dot\theta_n(t') + i l'\int_{t_0}^{t+\tau} dt'\theta_n(t')}\rangle_{\bs{\xi},\bs{\omega}} \\
%\label{eq:Cxi_b_deriv_ref}
%&=  \sum_l |A_l|^2  \langle e^{i l \omega_n \tau }\rangle_{\bs{\omega}}  
% \langle e^{i l \int_t^{t+\tau} dt'\xi_n(t') }\rangle_{\bs{\xi}} \\
%&= \sum_l |A_l|^2 \Phi(l\tau)\langle e^{il y(\tau;t)}\rangle \\
%&= \sum_l |A_l|^2 \Phi(l\tau) e^{-\frac{l^2}{2} \langle y(\tau;t)^2\rangle}\\
%&= \sum_l |A_l|^2 \Phi(l\tau) e^{-l^2\int_{0}^{\tau} dt'(\tau-t') C_\xi(t')}.
%\end{align}
%\end{subequations}
\begin{subequations}
\label{eq:vML_Cxi_deriv2}
\begin{align}
\frac{C_{\xi}(\tau)}{K^2/N} &=  \sum_n\sum_{l,l'} A_l A_{l'} \langle  e^{i l \theta_n(t) + il'\theta_n(t+\tau)}\rangle_{\bs{\xi},\bs{\omega},\bs{\theta_0}} \\
&= \sum_n \sum_{l,l'} A_l A_{l'} \langle e^{i(l+l')\theta_n(t_0)} \rangle_{\bs{\theta_0}} \times \nonumber \\
&\qquad \qquad \langle e^{i l \int_{t_0}^t dt'\dot\theta_n(t') + i l'\int_{t_0}^{t+\tau} dt'\theta_n(t')}\rangle_{\bs{\xi},\bs{\omega}} \\
\label{eq:Cxi_b_deriv_ref}
&=  \sum_n \sum_l |A_l|^2  \langle e^{i l \omega_n \tau }\rangle_{\bs{\omega}}  
 \langle e^{i l \int_t^{t+\tau} dt'\xi_n(t') }\rangle_{\bs{\xi}} \\
&= \sum_n\sum_l |A_l|^2 \Phi(l\tau)\langle e^{il y_n(\tau;t)}\rangle \\
&= \sum_n \sum_l |A_l|^2 \Phi(l\tau) e^{-\frac{l^2}{2} \langle y_n(\tau;t)^2\rangle}\\
&= N \sum_l |A_l|^2 \Phi(l\tau) e^{-l^2\int_{0}^{\tau} dt'(\tau-t') C_\xi(t')}.
\end{align}
\end{subequations}

A few remarks are in order: To ensure that the result does not depend on a particular initialization, one can average over the absolute phases $\theta_n(t_0)$ at a given (arbitrary) reference time $t_0$ and require that this average can be taken independently of the average over the $\xi_n$. For uniformly distributed phases $\theta_n(t_0)$, one immediately obtains $\langle e^{i(l+l')\theta_n(t_0)} \rangle_{\theta_n(t_0)} = \delta_{l,l'}$. We furthermore introduced the new variable $y_n(\tau;t) = \int_{t}^{t+\tau} dt' \xi_n(t')$. In the limit of a large network ($N\gg1$), one can assume that the $\xi_n$ are Gaussian distributed, and thus that $y_n(\tau;t)$ obeys the relation $\langle e^{ily_n}\rangle=e^{-l^{2}\langle y_n^2 \rangle/2}$. The second moment $\langle y_n^2\rangle$ can be found according to
\begin{align}
\langle y_n(\tau;t)^2\rangle &=  \int_{t}^{t+\tau} \!\!\! dt_1\int_{t}^{t+\tau}\!\!\! dt_{2} \left\langle \xi_{n}(t_{1}) \xi_{n}(t_{2})\right\rangle \nn \\
&= 2 \int_{0}^{\tau}\!\!\! dt (\tau-t) C_{\xi}(t) 
\end{align}
after a change of variables and using the symmetry of $C_{\xi}(t)$ around the origin. Since all units are statistically equivalent and their correlation functions the same, the sum over $n$ then becomes trivial.  
%(Note that the explicit $t$-dependence of $y(\tau;t)$ disappears after averaging.) 

The differential equation for $\Lambda$ (Eq.~\eqref{eq:vML-Lambda}) is eventually obtained with the substitution Eq.~\eqref{eq:Lambda}, and the noise autocorrelation follows as
\begin{equation}
\label{eq:vML-Cxi}
C_{\xi}(\tau) = K^2\sum_l |A_l|^2 \Phi(l\tau)\exp[-l^2\Lambda(\tau)].
\end{equation}

The autocorrelation function $C_{x_m}(t)$ of the pointer $x_m(t) = e^{i\theta_m(t)}$ for a given rotator $m$ is then rather straightforwardly obtained as 
\begin{align*}
\langle x_m(t)^*x_m(t+\tau)\rangle &= \langle  e^{i (\theta_m(t+\tau) - \theta_m(t))}\rangle \\
&= e^{i\omega_m\tau}  \langle  e^{i\int_{t}^{t+\tau} dt' \xi_m(t')}\rangle \\
&= e^{i\omega_m\tau - \Lambda(\tau)} .
\end{align*}
Of note, $C_{x_m}(t)$ explicitly depends on the particular intrinsic frequency $\omega_m$ of the rotator $m$. Averaged over all rotators, the autocorrelation becomes $C_x(t) = \Phi(\tau)e^{-\Lambda(\tau)}$.

\section{Rotator network with individual noise sources}
\label{sec:intrinsic}

Focussing on recurrent input, van Meegen \& Lindner did not address the effect of external inputs or other noise sources acting on individual units. Typically, however, in many applications, e.g.~in biological neural networks, the single units are subject to intrinsic noise, e.g.~channel noise or unreliable synaptic transmission in the case of neural networks~\cite{Tuckwell:1998,Koch:2004uc,Rusakov:2020dl}. 

It seems important to understand the degree to which these additional inputs may shape the dynamics of the network and individual units. There is a rich literature on the Kuramoto model driven by individual and common noise beginning with Kuramoto himself, see e.g.~\cite{Kuramoto:1984wo,Pikovsky:2015cs,Schafer:2018kk}, 
%% JR: didn't remember (not in your bib file) Sch15
but how noise affects the rotator dynamics considered here is not known. In this section, we will consider the case of intrinsic noise sources $\eta_m$ that are uncorrelated between different units (correlated inputs will be addressed in section~\ref{sec:common}). The model then reads
\begin{equation}
\label{eq:model-a}
\dot \theta_m = \omega_m + \sum_n K_{mn} f(\theta_n)  +\eta_m,
\end{equation}
the recurrent input again being specified by the connectivity matrix $K_{mn}$ and coupling function $f(\cdot)$. We consider the intrinsic fluctuations to be independent Gaussian noise sources with $\langle\eta_m\rangle=0$, $\langle\eta_m(t)\eta_n(t+\tau)\rangle=\delta_{mn}C_\eta(\tau)$. 

\subsection{Derivation of the self-consistent correlation functions in the presence of intrinsic noise}

We aim to determine how the presence of additional noise may shape the rotator autocorrelation function $C_{x_m}(\tau)$ of rotator $m$, the effective stochastic dynamics of which is given by 
\begin{equation}
\label{eq:model-a_bis}
\dot \theta_m = \omega_m + \xi_m + \eta_m,
\end{equation}
where the statistics of the network noise $\xi_m$ is still to be determined. 
% st $\langle\xi_m(t)\xi_m(t')\rangle=C_\xi(t-t')$ is
First of all, it is not difficult to show that the network noise is not correlated with the external noise, as seen as follows: 
\begin{align}
\label{eq:K_single_power}
\langle \xi_m(t)\eta_m(t')\rangle &= \sum_n \langle K_{mn}\rangle_K \langle f(\theta_n(t)) \eta_m(t')\rangle = 0
\end{align}
since $\langle K_{mn}\rangle_K=0$. 

We now turn to the problem of calculating the autocorrelation function of the network noise. When all rotators are subject to intrinsic noise, we find a modified self-consistency condition for $C_\xi(t)$ as follows. Along the lines of the derivation in the case without noise, Eq.~\eqref{eq:Cxi_b_deriv_ref} now becomes
%\begin{subequations}
\begin{align}
\nonumber
\frac{C_\xi(\tau)}{K^2/N} &= \sum_{n}\sum_l |A_l|^2  \langle e^{i l \omega_n \tau }\rangle_{\bs{\omega}}
 \langle e^{i l \int_t^{t+\tau} dt'[\xi_n(t')+\eta_n(t')]  }\rangle_{\bs{\xi},\bs{\eta}} \\
&= \sum_{n}\sum_l |A_l|^2 \Phi(l\tau)\langle e^{il y_{n}(\tau;t)}\rangle,
\end{align}
%\end{subequations}
where we redefined the variable 
\begin{equation}
\label{eq:y_w_intrinsic}
y_{n}(\tau;t) = \int_t^{t+\tau}dt' [\xi_n(t')+\eta_n(t')] .
\end{equation}
Assuming that $y_{n}(\tau;t)$ is Gaussian, we now find
\begin{subequations}
\label{eq:y_calc}
\begin{align}
\langle e^{i l y_{n}(\tau)} \rangle &=  e^{-\frac{l^2}{2} \langle y_{n}(\tau)^2\rangle} \\
&=  e^{- l^2 \int_0^\tau dt (\tau-t)  \langle  [\xi_n(t')+\eta_n(t')] [\xi_n(t'+t)+\eta_n(t'+t)]\rangle }\\
&=  e^{- l^2 \int_0^\tau dt (\tau-t) [C_\xi(t) + C_\eta(t)]}.
\end{align}
\end{subequations}
Putting everything together, one obtains 
\begin{align}
\label{eq:a-Cxi}
C_\xi(\tau) &= K^2\sum_l |A_l|^2 \Phi(l\tau)\exp\!{\(-l^2 [\Lambda(\tau) + \Lambda_\eta(\tau)]\)}.
\end{align}
Here $\Lambda_\eta(\tau) = \int_0^\tau dt(\tau-t)C_\eta(t)$ is given by the autocorrelation function $C_\eta(t)$ of the intrinsic noise 
%sources\footnote{Note that the integral $\int_0^\tau dt(\tau-t)C_\eta(t)$ can be either analytically calculated for simple processes or determined by numerically integrating the ordinary differential equation for $\Lambda_\eta(\tau) = \int_0^\tau dt(\tau-t)C_\eta(t)$ with $\ddot\Lambda_\eta(t) = C_\eta(t)$, $\Lambda_\eta(0)=0$, $\dot\Lambda_\eta(0)=0$.} 
sources\footnote{Note that the integral can be either analytically calculated for simple processes or determined by numerically integrating the ordinary differential equation for $\Lambda_\eta(\tau)$ with $\ddot\Lambda_\eta(t) = C_\eta(t)$, $\Lambda_\eta(0)=0$, $\dot\Lambda_\eta(0)=0$.} 
and the function $\Lambda(\tau) = \int_0^\tau dt(\tau-t)C_\xi(t)$, defined as before, now obeys the ordinary differential equation
\begin{equation}
\label{eq:a-Lambda}
\ddot\Lambda(\tau) = K^2\sum_l |A_l|^2 \Phi(l\tau)\exp\!{\(-l^2 [\Lambda(\tau) + \Lambda_\eta(\tau)]\)}.
\end{equation}

Once we have calculated $\Lambda(t)$, we know the autocorrelation function of the network noise $C_\xi(t)$ via Eq.~\eqref{eq:a-Cxi}, but can now also determine the autocorrelation function $C_{x_m}(t)$ of the $m$-th oscillator (the complex pointer $x_m=e^{i\theta_m}$):  
\begin{equation}
\label{eq:Cx_a}
C_{x_m}(\tau) = \exp\!\left[i\omega_m\tau - \Lambda(\tau) - \Lambda_\eta(\tau)\right].
\end{equation}
Similarly to the case without intrinsic noise, this can be derived using $\theta_m(t) = \theta_m(t_0) + \int_{t_0}^t dt'\dot\theta_m(t')$ with Eq.~\eqref{eq:model-a_bis} and using Eq.~\eqref{eq:y_calc} for $l=1$. Eqs.~(\ref{eq:a-Cxi}-\ref{eq:Cx_a}) constitute the theory for the self-consistent correlation functions for a rotator network with individual intrinsic Gaussian noise.

So far we have not made any assumptions about the temporal correlations of the intrinsic noise, i.e., our theory holds for a general colored noise $\eta_m(t)$. For our numerical examples in the following, we consider for simplicity Gaussian white noise, for which $C_\eta(\tau)=2D_\eta \delta(\tau)$, implying $\Lambda_\eta(\tau)=D_\eta \tau$. 

%One might be tempted to think that the sole effect of the intrinsic noise on $C_{x_m}(t)$ is the exponential damping with a damping time constant $1/D_\eta$ in Eq.~\eqref{eq:Cx_a}. This relies on the tacit assumption that the network noise correlation function $C_\xi(t)$ would still be given in terms of $\Lambda(t)$ for the recurrent rotator network without intrinsic noise sources, determined by Eq.~\eqref{eq:vML-Lambda}. 

\subsection{Effects of intrinsic white noise on the rotator autocorrelation}

\begin{figure}[tb]
\includegraphics[width=0.8\columnwidth]{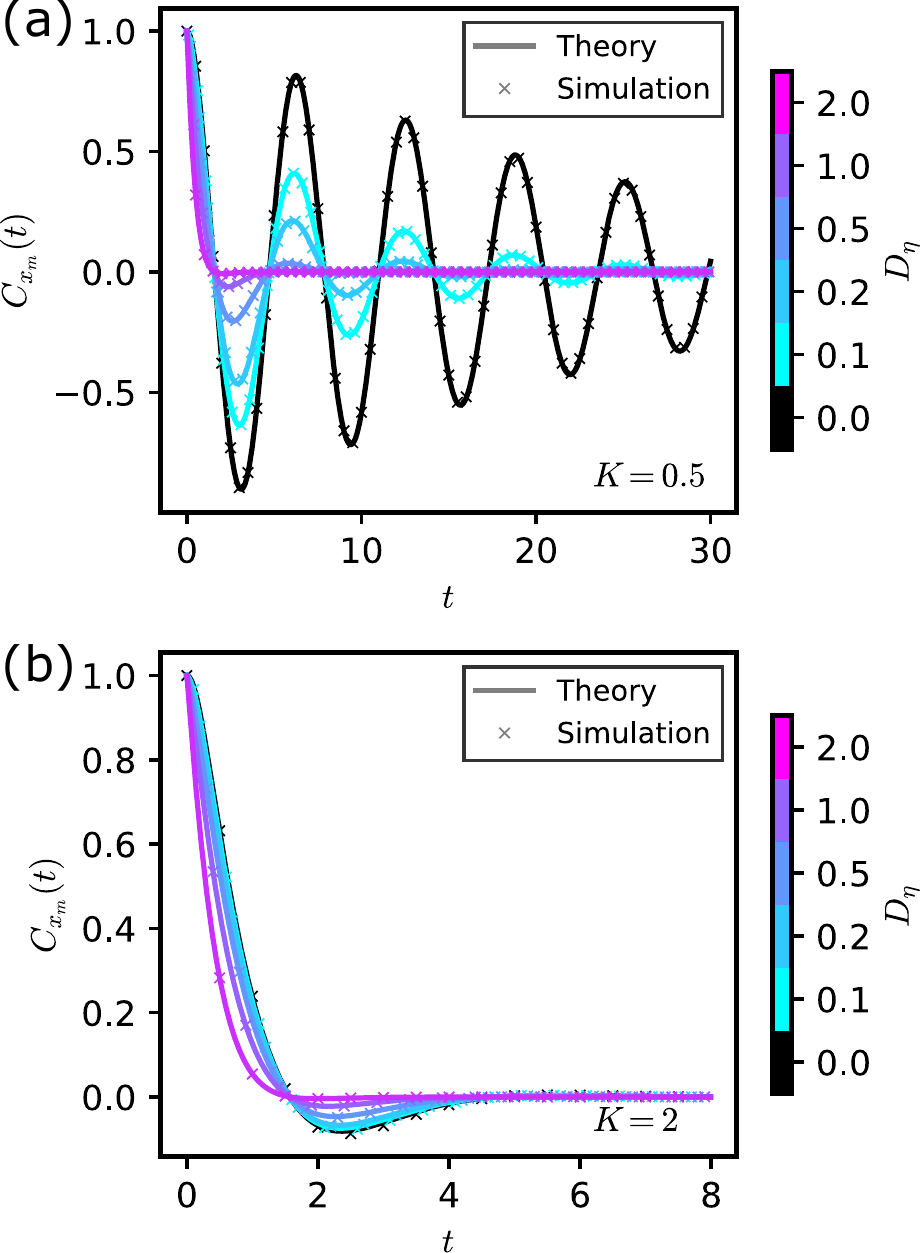}
\label{fig:ind_noise_Cx}
\caption{Rotator autocorrelation function $C_{x_m}$ in the presence of intrinsic noise, for weak (a) and strong (b) recurrent connection strengths. Theoretical results (solid lines) and stochastic network simulations (crosses) are shown for coupling function $f(\theta)=\sin(2\theta) + \cos(3\theta)$ and parameters $\omega_0=1$, $\sigma_\omega=0$, and (a) $K=0.5$, respectively (b) $K=2$. The size of the simulated networks was $N=100$. 
Simulations were carried out with a time step $dt=0.01$, stationary initial conditions (i.e.~uniformly distributed phases), for a total duration of $T=25 T_0$ with $T_0=2500$, and $10$ different realizations of $K_{mn}$. Statistics were computed for bouts of length $T_0$ of the phase trajectories, and averages subsequently taken over bouts, network realizations, and rotators. The theory was evaluated with a time step $dt=0.001$.  
\label{fig:ind_noise_Cx}}
\end{figure}

In order to test our theory, we simulated the rotator network with additional intrinsic white noise for various combinations of noise intensity $D_\eta$ and coupling strength $K$. The corresponding theoretical predictions for $C_{x_m}(t)$ obtained from Eqs.~\eqref{eq:a-Lambda} and~\eqref{eq:Cx_a} are nicely confirmed for all combinations used, see Fig.~\ref{fig:ind_noise_Cx}, although the number of units in the network ($N=100$) is not excessively large. For a white noise, Eq.~\eqref{eq:Cx_a} reads 
\begin{equation}
C_{x_m}(\tau) = \exp\!\left[i\omega_m\tau - \Lambda(\tau) - D_\eta \tau \right],
\end{equation}
from which we can expect a damping of the autocorrelation function with increasing noise strength $D_\eta$. This is indeed observed in the case of both weak (Fig.~\ref{fig:ind_noise_Cx}a) and strong (Fig.~\ref{fig:ind_noise_Cx}b) network noise. 

The network-mediated influence of the intrinsic noises $\eta_n$, $n\neq m$, on the autocorrelation function of rotator $m$ enters the expression for $C_{x_m}(t)$ via the modified network-noise autocorrelation function $C_\xi(t)$. In order to assess this indirect effect of the intrinsic noise, we compare our result to the case where only a specific, single unit $m$ receives additional intrinsic noise $\eta_m$ according to Eq.~\eqref{eq:model-a}, while all other units of the network evolve according to Eq.~\eqref{eq:vML-model}. For the particular rotator $m$ that is subject to intrinsic  noise, $C_{x_m}(t)$ is then still given by Eq.~\eqref{eq:Cx_a} with the exponential damping term $e^{-D_\eta t}$, but where the network-noise autocorrelation $C_\xi(t)$ is the ``intrinsic-noise free'' version $C_\xi(t)=\ddot\Lambda(t)$ with $\Lambda$ obeying Eq.~\eqref{eq:vML-Lambda}. 

\begin{figure}[tb]
{\includegraphics[width=0.8\columnwidth]{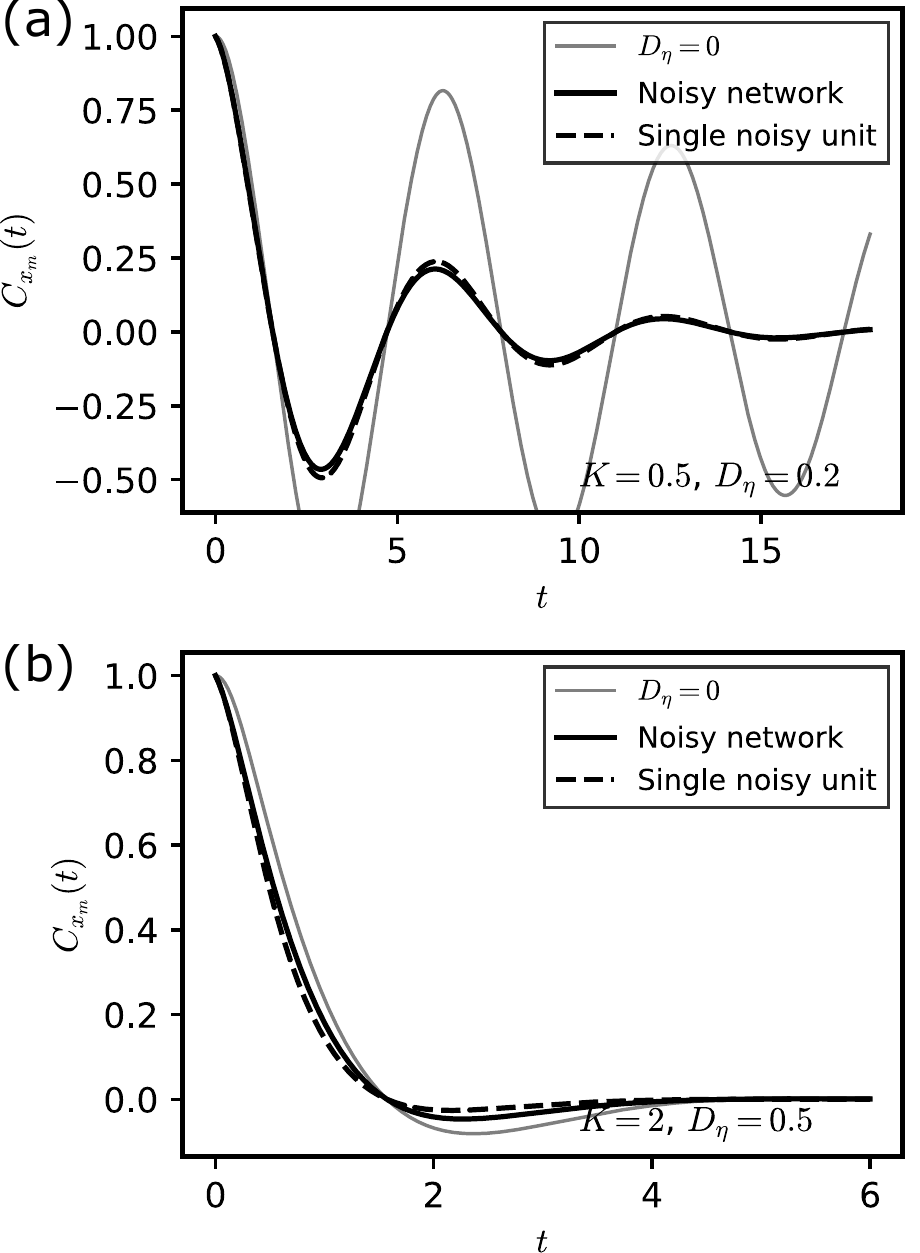}}
\label{fig:ind_noise_Cx_compare}
\caption{The network-mediated effect of noise in ``background'' units on rotator autocorrelation function $C_{x_m}$ of noisy unit $m$ (solid lines) can be assessed by comparison with the autocorrelation function $C_{x_m}$ of unit $m$ in the case where only unit $m$ is subject to noise (dashed lines). The autocorrelation in the absence of any noise is also shown (thin lines). (a) Intermediate noise level with $K=0.5$, $D_\eta=0.2$; (b) strong noise level with $K=2$, $D_\eta=0.5$. All curves are obtained from the theory, with $dt=0.001$.
\label{fig:ind_noise_Cx_compare}}
\end{figure}

Stochastic network simulations and theoretical predictions for $C_x(t)$ again nicely agree (not shown) but look very similar to the case where all units are subject to intrinsic noise shown in Fig.~\ref{fig:ind_noise_Cx}. We nevertheless identify two distinct effects due to the network-mediated influence of the other intrinsic noise sources, as shown in Fig.~\ref{fig:ind_noise_Cx_compare}: For weak network noise (Fig.~\ref{fig:ind_noise_Cx_compare}a), we find that the intrinsic noise on the other units in the network further decorrelates oscillator $m$ relative to the case of a single noisy unit with ``intrisic-noise free'' recurrent input. Interestingly, in the case of strong network noise (Fig.~\ref{fig:ind_noise_Cx_compare}b), the opposite effect can be observed, as the autocorrelation of the specific rotator $m$ decays faster in the case where it is the only one subject to intrinsic noise while the rest of the network does evolve without additional intrinsic noise.

%\begin{figure}[h!]
%\caption{(a) Sketch of recurrent network, (b) phase variables, (c,d) noise power spectra with vs. without correlated noise: (c) Gaussian network c=0,1, (d) Dale with p=0.1, 0.5}
%\end{figure}
%
%\begin{figure}[h!]
%\caption{Heat map of skew and kurtosis as a function of parameters for c=1.}
%\end{figure}
%
%
%\begin{figure}[h!]
%\caption{Rescaled cumulants, correlation functions, and power spectra for two parameter set 1.}
%\end{figure}
%
%
%\begin{figure}[h!]
%\caption{Rescaled cumulants, correlation functions, and power spectra for two parameter set 2.}
%\end{figure}

%\section{Rotator network with common external input inducing non-Gaussian statistics}
\section{Rotator network with common external input}
\label{sec:common}

So far we extended the rotator network model of Ref.~\onlinecite{vanMeegen:2018hj} to the case where individual units are subject to intrinsic noise, and assumed independence of the noise sources for different units. In the context of neural networks, possible origins of such a noise term could be the neurons' stochastic ion channel kinetics and spontaneous release of neurotransmitters~\cite{Tuckwell:1998}, for which the assumption of independence seems clearly satisfied. Another possible interpretation of the noise term $\eta_{m}$ is that it results from external (as opposed to recurrent) synaptic inputs that are uncorrelated across units. %However, this latter assumption of statistical independence of inputs from an external population is rarely satisfied. 

More generally however, neurons may also receive \emph{correlated} external inputs, which could stem e.g.~from other brain areas that project broadly to a local recurrently connected network. Here, we will thus relax the assumption that any additional inputs to the individual units are independent and study the potential effects on the network dynamics. Regarding the theory, one might be tempted to think that it can be extended to this new case without major changes and that its main assumptions, e.g.~the Gaussian distribution of the network noise, are still valid.
%While it is not obvious \emph{a priori} whether the theory can be extended to this case without major changes. 
We will show however that 
%even if the external noise is Gaussian, 
the network noise ceases to be Gaussian-distributed if the external noise is significantly correlated across units. 
%We hence face the peculiar situation where the addition of external Gaussian noise turns a previously Gaussian into a non-Gaussian network noise. 
Consequently, the self-consistent theory of Ref.~\onlinecite{vanMeegen:2018hj} cannot easily be extended to this case. Instead, we have to more substantially rework the theory. By taking higher-order correlation functions of the network noise into account, we can arrive at an approximate theory that captures the resulting non-Gaussian network statistics surprisingly well. 

To be specific, we consider the following model, where $\etac$ is the additional input common to all units:
\begin{equation}
\label{eq:model-c}
\dot \theta_m = \omega_m + \sum_n K_{mn} f(\theta_n)  +\eta_m +\etac.
\end{equation}
We assume that $\etac$ is a Gaussian noise which is uncorrelated with the intrinsic noise $\eta_m$ of each unit $m$, $\langle\etac(t)\eta_m(t+\tau)\rangle\equiv0$, and that its temporal statistics are described by the autocorrelation function $\langle\etac(t)\etac(t+\tau)\rangle=\Cc(\tau)$. Furthermore, along the same lines as sketched for the intrinsic noise $\eta_m$,  Eq.~\eqref{eq:K_single_power}, we note that $\etac$ is uncorrelated with the network noise $\xi_{m}$, $\langle\etac(t)\xi_m(t+\tau)\rangle=0$, by virtue of the averaging properties of $K_{mn}$. 

%We can try to proceed along the same lines as in the previous cases without noise (Sec.~\ref{sec:previous}) and with independent stochastic forcing (Sec.~\ref{sec:intrinsic}) 
%With the aim to derive a self-consistent equation for the network-noise autocorrelation function $C_\xi(\tau)$, 

To develop the theory for the case with common input, we start again by expressing the autocorrelation $\langle\xi_m(t)\xi_m(t+\tau)\rangle$ in terms of the connectivity matrix $K_{mn}$, coupling function $f(\cdot)$, and rotator phases $\theta_m$. 
%\begin{align*}
%C_{\xi}(\tau) &= \sum_{n,n'} \langle K_{mn}K_{mn'} f(\theta_n(t))f(\theta_{n'}(t+\tau))\rangle_{K,\bs{\omega},\bs{\theta_0},\bs{\eta},\etac}.
%\end{align*}
The additional average over independent realizations of the common noise $\etac$ is the only departure from the case considered in the previous section and does not interfere with the initial steps of Eqs.~\eqref{eq:vML_Cxi_deriv1} and~\eqref{eq:vML_Cxi_deriv2}, which now lead to 
\begin{multline}
\label{eq:Cxi_c_1}
C_\xi(\tau) = \frac{K^2}{N}\sum_{N} \sum_l |A_l|^2  \langle e^{i l \omega_n \tau }\rangle_{\bs{\omega}}  \times \\
 \langle e^{i l \int_t^{t+\tau} dt'[\xi_n(t')+\eta_n(t')+\etac(t')]  }\rangle_{\bs{\xi},\bs{\eta},\etac} .
\end{multline}
By redefining the variable $y_{n}(\tau;t)$ according to 
\begin{equation}
\label{eq:y_def_common}
y_{n}(\tau;t) = \int_t^{t+\tau}dt' [\xi_n(t')+\eta_n(t') + \etac(t')] 
\end{equation}
and noting that $\langle e^{i l \omega_n \tau }\rangle_{\bs{\omega}}= \Phi(l \tau)$, expression~\eqref{eq:Cxi_c_1} can again be written more conveniently as 
\begin{equation}
\label{eq:Cxi_c_2}
C_\xi(\tau) = \frac{K^2}{N} \sum_n \sum_l |A_l|^2  \Phi(l \tau) \langle e^{i l y_n(\tau;t)}\rangle_{\bs{\xi},\bs{\eta},\etac} .
\end{equation}

A naÃ¯ve generalization of our previous approach would yield a simple modification of Eqs.~(\ref{eq:a-Cxi}-\ref{eq:Cx_a}). Under the assumption that the $y_n(\tau;t)$ were still Gaussian-distributed, it would follow that the respective expressions hold with the autocorrelation function $C_\eta(\tau)$ of the intrinsic noise simply being replaced by the sum of the noise autocorrelation functions $C_\eta(\tau)+\Cc(\tau)$. Consider the special situation in which intrinsic noise and common noise share the same autocorrelation function and we can describe by a parameter $c$ how much of the input noise is common, i.e.~$C_{\eta}(\tau)=(1-c)C_{\rm tot}(\tau)$ and $\Cc(\tau)=c C_{\rm tot}(\tau)$. In this case, the output statistics would not depend on $c$, implying that the network statistics is unaffected by the presence or absence of correlations among the additional stochastic forcings of the rotators. However, this lack of dependence is \emph{not} observed in stochastic network simulations of Eq.~\eqref{eq:model-c}, demonstrating that the naÃ¯ve generalization sketched above indeed fails to account for the effective stochastic dynamics of the system.

In order to derive a correct self-consistent expression for the network-noise autocorrelation function $C_{\xi}(\tau)$, we thus need to determine the only remaining unknown term $\langle e^{i l y_{n}(\tau;t)}\rangle_{\bs{\xi},\bs{\eta},\etac}$ in Eq.~\eqref{eq:Cxi_c_2}. Let us consider the general case of a stochastic variable $y$. By definition of the cumulant-generating function, we have 
\begin{equation}
\label{eq:y_cumulants}
\langle e^{i l y}\rangle_y = \exp\(\sum_{k=1}^{\infty}\frac{(il)^{k}}{k!}\kappa_{k}\).
\end{equation}
Here, the $\kappa_{k}$ are the cumulants of $y$ which we can calculate as a function of its moments $\langle y^{j} \rangle_{y}$, $j\le k$. If $y$ is zero-centered and Gaussian-distributed, one immediately finds $\langle e^{i l y}\rangle_y = e^{-l^2\kappa_2/2}=e^{-l^2\langle y^2\rangle_{y}/2}$, which led to the results obtained in the previous sections. We will see that, while $y_{n}(\tau;t)$ as defined  in Eq.~\eqref{eq:y_def_common} still has a vanishing mean value in the presence of common input, it is no longer Gaussian-distributed and higher cumulants need to be taken into account. 

Assuming that the deviations from Gaussian statistics are small, our strategy is the following: First, we approximate the average $\lr{e^{i l y}}_y$ by taking only the first few cumulants according to Eq.~\eqref{eq:y_cumulants} into account, i.e.,
\begin{equation}
\label{eq:y_cumulants_approx}
\langle e^{i l y}\rangle_y \approx \exp\!\(-l^2\frac{\kappa_2}{2} -il^3\frac{\kappa_3}{6} + l^4\frac{\kappa_4}{24}\).
\end{equation}
Here, the higher-order cumulants $\kappa_3$ and $\kappa_4$ are finite only in the presence of correlations between the inputs of different units, and should vanish with vanishing correlations. As we will show, these higher-order cumulants have in turn themselves Gaussian and non-Gaussian contributions in a sense that will be made explicit below. In line with our assumption of a small departure from Gaussian statistics, we thus, in a second approximation, calculate each of these cumulants to zeroth order, that is, assuming purely Gaussian statistics of composite variables of the type $y_n+\etac$ and $y_n+y_m$. This strategy allows us to obtain a self-consistent equation for $C_\xi(\tau)$ in the presence of common input that reproduces measured autocorrelation functions as well as higher-order cumulants of $y_n$ in stochastic network simulations. 

\subsection{Calculation of the cumulants of $y_n(\tau;t)$}

The first cumulant is identical to the first moment and vanishes, 
\begin{equation}
\langle y_{n}(\tau;t)\rangle = \intl_t^{t+\tau}dt' [\lr{\xi_n(t')}+\lr{\eta_n(t')} + \lr{\etac(t')}] =0,
\end{equation}
as $\lr{\xi_{n}}=0$ because of the average over the $K_{nn'}$ and $\lr{\eta_{n}}=\lr{\etac}=0$ by definition. If the first moment vanishes, the second cumulant $\kappa_2$, i.e.~the variance, is identical to the second moment. Its calculation is analogous to the case of purely intrinsic noise discussed above, and given by
\begin{equation}
\label{eq:kappa2_common}
\langle y_{n}(\tau)^{2}\rangle = \intl_0^\tau dt (\tau-t) [C_\xi(t) + C_\eta(t) + \Cc(t)].
\end{equation}
The new non-trivial contributions in this theory will be the third and fourth cumulants which can also be expressed in terms of the moments; in the following we will calculate and use 
\begin{align*}
\kappa_{3} &=  \lr{y_{n}^{3}} , \\
\kappa_{4} &=  \lr{y_{n}^{4}} - 3\lr{y_{n}^{2}}^{2},
\end{align*}
to approximate the characteristic function in Eq.~\eqref{eq:Cxi_c_2}.

\subsubsection{Third cumulant}

Inserting the definition of $y_n$, we obtain for the third cumulant 
\begin{align}
\nonumber
\kappa_{3}(\tau) &= \langle y_{n}^{3}(\tau;t)\rangle \\
%\nonumber
%&= \left\langle \prod_{j=1}^{3}\int_{t}^{t+\tau}\!\!\! dt_{j}[\xi_{m}(t_{j})+\eta_{m}(t_{j})+\etac(t_{j})]\right\rangle \\
\label{eq:kappa3_1}
&= \left\langle \prod_{j=1}^{3} \intl_{t}^{t+\tau} \!\!\! dt_j [\xi_{n}(t_{j})+\eta_{n}(t_{j})+\etac(t_{j})]\right\rangle . %\\
%\label{eq:kappa3_1}
%&= \intl_{t}^{t+\tau} \!\!\! dt_1\intl_{t}^{t+\tau}\!\!\! dt_{2}\intl_{t}^{t+\tau} \!\!\! dt_{3} \left\langle \prod_{j=1}^{3}[\xi_{m}(t_{j})+\eta_{m}(t_{j})+\etac(t_{j})]\right\rangle.
\end{align}
Note  that $\kappa_{3}$ does neither depend on $t$ nor on any neuron index $n$, as these dependences disappear after averaging over uniform initial conditions for the phases and the disorder of the connectivity, respectively. 

Before evaluating the triple integral, we consider separately the different contributions that arise from the correlator of the different noise sources, such as~$\lr{\xi_{n}(t_{1})\eta_{n}(t_{2})\etac(t_{3})}$ or $\lr{\xi_{n}(t_{1})\xi_{n}(t_{2})\etac(t_{3})}$. Because $\eta_{n}$ and $\etac$ are Gaussian and uncorrelated by construction, all terms in which only these two appear vanish. Furthermore, averaging over the zero-centered, Gaussian $K_{mn}$ implies that all terms with odd powers of $\xi_{n}$ must also vanish, see e.g.~Eq.~\eqref{eq:K_single_power}. As all three integrals in Eq.~\eqref{eq:kappa3_1} cover the same domain, we can arbitrarily permute the variables $t_j$, and retain the following remaining contributions:
\begin{multline}
\label{eq:kappa3_2}
%\left\langle \prod_{j=1}^{3}[\xi_{m}(t_{j})+\eta_{m}(t_{j})+\etac(t_{j})]\right\rangle/3 \stackrel{\wedge}{=}  \\
%\langle\xi_m(t_1)\xi_m(t_2)\eta_m(t_3)\rangle + \langle\xi_m(t_1)\xi_m(t_2)\etac(t_3)\rangle 
\big\langle \prod_{j=1}^{3}[\xi_{n}(t_{j})+\eta_{n}(t_{j})+\etac(t_{j})]\big\rangle/3 = \\ 
\langle\xi_n(t_1)\xi_n(t_2)\eta_n(t_3)\rangle 
+ \langle\xi_n(t_1)\xi_n(t_2)\etac(t_3)\rangle ,
\end{multline}
where the equality holds under the triple integral of Eq.~\eqref{eq:kappa3_1}.

We next aim to express both contributions in terms of the correlation functions $C_{\xi}$, $C_{\eta}$, and $\Cc$, where the latter two are known and $C_{\xi}$ remains to be determined self-consistently. To be specific and in line with our treatment of $\eta_m$ in the previous section, we will consider both $\eta_{m}$ and $\etac$ to be Gaussian white noises with respective intensities $D_\eta$ and $\Dc$ for the remainder of this section. The correlation functions are then given by $C_{\eta}(\tau)=2D_{\eta}\delta(\tau)$,  $\Cc(\tau)=2\Dc\delta(\tau)$, which decidedly simplifies calculations. Using the definition of $\xi_{m}$, we can expand the first contribution of Eq.~\eqref{eq:kappa3_2} as
\begin{align}
&\mkern-36mu \langle\xi_m(t_1)\xi_m(t_2)\eta_m(t_3)\rangle \nn \\
&= \sum_{n,n'}\langle K_{mn}K_{mn'}f(\theta_n(t_1))f(\theta_{n'}(t_2))\eta_m(t_3)\rangle \nn \\
&= \sum_{n,n'}\langle K_{mn}K_{mn'}\rangle_{K} \langle f(\theta_n(t_1))f(\theta_{n'}(t_2))\eta_m(t_3)\rangle \nn \\
&= \frac{K^2}{N}\sum_{n} \langle f(\theta_n(t_1))f(\theta_n(t_2))\eta_m(t_3)\rangle .
\end{align}
Again along the lines of Eqs.~\eqref{eq:vML_Cxi_deriv1} and ~\eqref{eq:vML_Cxi_deriv2} but with $\dot \theta_m = \omega_m + \xi_m  +\eta_m +\etac$, we now obtain 
\begin{multline}
\label{eq:k3_deriv_xi2etam_1}
\langle\xi_m(t_1)\xi_m(t_2)\eta_m(t_3)\rangle = \frac{K^2}{N} \sum_n \sum_l |A_l|^2  \Phi(l (t_2-t_1)) \times \\
\langle e^{i l y_n(t_2-t_1;t_1)} \eta_m(t_3)\rangle_{\bs{\xi},\bs{\eta},\etac} .
\end{multline}
Instead of the characteristic function $\langle e^{i l y_n}\rangle$, we now need to find an expression for the average $\langle e^{i l y_n}\eta_m\rangle$. 

We can distinguish the cases $n\neq m$ and $n=m$. When $n\neq m$, $\langle y_{n}(t_2-t_1) \eta_{m}(t_3)\rangle=0$, since $\xi_n$, $\eta_n$, and $\etac$ are all uncorrelated with $\eta_m$; therefore, $\langle e^{i l y_n(t_2-t_1;t_1)} \eta_m(t_3)\rangle = \lr{e^{i l y_n(t_2-t_1;t_1)}}\lr{\eta_m(t_3)} = 0$. When $n=m$, the correlation $\langle e^{i l y_n(t_2-t_1;t_1)} \eta_n(t_3)\rangle$ does not vanish but should remain bounded for large $N$. We thus expect that, according to Eq.~\eqref{eq:k3_deriv_xi2etam_1}, $\langle\xi_n(t_1)\xi_n(t_2)\eta_n(t_3)\rangle$ scales like $1/N$ and vanishes for $N\to\infty$. Hence we are allowed to neglect the first term on the r.h.s. of Eq.~\eqref{eq:kappa3_2}.

In order to calculate the last remaining term $\langle\xi_n(t_1)\xi_n(t_2)\etac(t_3)\rangle$ of Eq.~\eqref{eq:kappa3_2}, we need to find the average $\langle e^{i l y_n}\etac\rangle$ in analogy to Eq.~\eqref{eq:k3_deriv_xi2etam_1} for $\etac$. We exploit the following relation for stochastic variables $a$ and $b$:
%\begin{align}
%\langle e^{a} b \rangle &= \left\langle \( -i \frac{d}{ds} e^{i(ra+sb)}\)_{r=1,s=0}\right\rangle \nn \\
%&= -i \( \frac{d}{ds}  \left\langle e^{i(ra+sb)}\right\rangle \)_{r=1,s=0} ,
%\end{align}
\begin{align}
\label{eq:exp_ab_rel1}
\langle e^{ia} b \rangle &= -i \( \frac{d}{dr}  \left\langle e^{i(a+rb)}\right\rangle \)_{r=0} .
\end{align}
%with  $r,s \in \mathbb{R}$. 
%The expectation $\langle e^{ia} b \rangle$ is expressed as the derivative of a characteristic function $\langle e^{iuz}\rangle_z$ evaluated at $u=1$, where $z=ra+sb$. 
In our case, $a=ly_n(t_2-t_1)$, $b=\etac(t_3)$. In Eq.~\eqref{eq:exp_ab_rel1}, the expectation $\langle e^{ia} b \rangle$ is expressed as the derivative of a characteristic function $\langle e^{iuz}\rangle_z$ evaluated at $u=1$, where $z=a+rb$. This characteristic function can again be expressed in terms of the cumulants of $z$, 
\begin{equation}
\label{eq:z_cumulants}
\langle e^{i u z}\rangle_z = \exp\(\sum_{k=1}^{\infty}\frac{(iu)^{k}}{k!}\hat\kappa_{k}\),
\end{equation}
where $\hat\kappa_k$ are the cumulants of $z$. The first cumulant $\hat\kappa_1$ of $z$ vanishes. The second cumulant can be calculated from the variances and covariances of $y_n$ and $\etac$. Above, we assumed that $y_n$ is (weakly) non-Gaussian in the presence of common input, and thus in turn $z$ cannot be Gaussian either. However, we  make here a Gaussian approximation and neglect higher-order cumulants of $z$, as these appear as the corrections to the corrections in Eq.~\eqref{eq:y_cumulants_approx}. We then find
\begin{equation}
\langle e^{i u z}\rangle_z \approx \exp\!\(-\frac{u^2}{2}\(\lr{a^2} + 2r\lr{ab} + r^2\lr{b^2}\) \),
\end{equation}
and consequently for the required expression
\begin{align}
\langle e^{ily_n} \etac \rangle &\approx i l \lr{y_{n}\etac}  e^{-l^2\lr{y_n^2}/2}  .
\end{align}
This term can now be inserted in the analogous version of Eq.~\eqref{eq:k3_deriv_xi2etam_1} for $\etac$, 
\begin{align}
\label{eq:k3_deriv_xi2etac_1}
&\mkern-24mu\langle\xi_m(t_1)\xi_m(t_2)\etac(t_3)\rangle \nn\\
&= \frac{K^2}{N} \sum_n \sum_l |A_l|^2  \Phi(l (t_2-t_1)) \times \nn\\
&\quad\quad \langle e^{i l y_n(t_2-t_1;t_1)} \etac(t_3)\rangle_{\bs{\xi},\bs{\eta},\etac}  \nn \\
&\approx  K^2 \sum_l |A_l|^2 i l \Phi(l (t_2-t_1))  \lr{y_n(t_2-t_1)\etac(t_3)} \times \nn \\
&\qquad\qquad e^{-l^2\lr{y_n(t_2-t_1)^2}/2} .
\end{align}
Here, we furthermore used that $\langle e^{ily_n} \etac \rangle$ does not depend on neuron index $n$ and therefore all $n$ terms of the sum contribute equally. The covariance of $y_n$ and $\etac$ can be calculated directly using the definition of $y_n$ in Eq.~\eqref{eq:y_def_common} and is given by
\begin{align}
\lr{y_{n}(t_2-t_1)\etac(t_3)} = 2 \Dc \Theta(t_3-t_1) \Theta(t_2-t_3).
\end{align}

We can now evaluate the triple integral of Eq.~\eqref{eq:kappa3_1} to calculate the third cumulant $\kappa_3(\tau)$, which based on the above can be expressed as
\begin{align}
\kappa_{3}(\tau) &\approx  12 \Dc K^ 2 \sum_l |A_l|^2 i l  \intl_{t}^{t+\tau} \!\!\! dt_1\!\! \intl_{0}^{t+\tau-t_1}\!\!\! d\tau' \tau' \times \nn\\
&\qquad \Phi(l \tau') \exp(-l^2\lr{y_n(\tau')^2}/2).
\end{align}	
Here, we furthermore used that $t_1$ and $t_2$ can be permuted to constrain the integral over $t_2$ to $t_2\ge t_1$, which gives an additional factor 2, and introduced the variable $\tau'=t_2-t_1$. By changing the order of integration, we can eliminate the integral over $t_1$, and using also the result for the second moment of Eq.~\eqref{eq:kappa2_common} with the by now familiar definition $\Lambda(\tau) = \int_0^\tau dt(\tau-t)C_\xi(t)$, we eventually obtain
\begin{align}
\label{eq:kappa3_result}
\kappa_{3}(\tau) &\approx  12 \Dc K^ 2 \sum_l |A_l|^2 i l  \intl_0^\tau \!\! d\tau' (\tau-\tau') \tau' \times \nn\\
&\qquad \Phi(l \tau') \exp(-l^2[\Lambda(\tau') + (D_\eta + \Dc)\tau']).
\end{align}	

In order to facilitate the numerical calculations and to avoid evaluating the integral over $\tau'$, we follow a similar strategy as in the cases without common noise, and compute the second time derivative of $\kappa_3(\tau)$,
\begin{align}
\label{eq:ddot_kappa3}
\ddot \kappa_{3}(\tau) &\approx  12 \Dc K^ 2 \sum_l |A_l|^2 i l  \tau \Phi(l \tau) e^{-l^2[\Lambda(\tau) + (D_\eta + \Dc)\tau]},
\end{align}	
which can be used to calculate $\kappa_3(\tau)$ along with $\Lambda(\tau)$ by integrating numerically the coupled ordinary differential equations for both quantities. To be precise, if we stopped our approximation for $\lr{e^{ily_n}}$ at the third cumulant, 
Eq.~\eqref{eq:Cxi_c_2} would become 
\begin{align}
\label{eq:ddot_Lambda_3cumul_only}
\ddot \Lambda(\tau) &\approx  K^ 2 \sum_l |A_l|^2  \Phi(l \tau) e^{-l^2[\Lambda(\tau) + (D_\eta + \Dc)\tau] - \frac{il^3}{6}\kappa_3(\tau)}.
\end{align}	
Eqs.~\eqref{eq:ddot_kappa3} and~\eqref{eq:ddot_Lambda_3cumul_only} together with the initial conditions $\kappa_3(0)=\dot\kappa_3(0)=\Lambda(0)=\dot\Lambda(0)=0$ (found from Eqs.~\eqref{eq:kappa3_result} and~\eqref{eq:Lambda})  would then constitute the self-consistent theory for the autocorrelation function of the network noise  $C_\xi(\tau) = \ddot\Lambda(\tau)$ up to third order.

\begin{figure*}
\includegraphics[width=\textwidth]{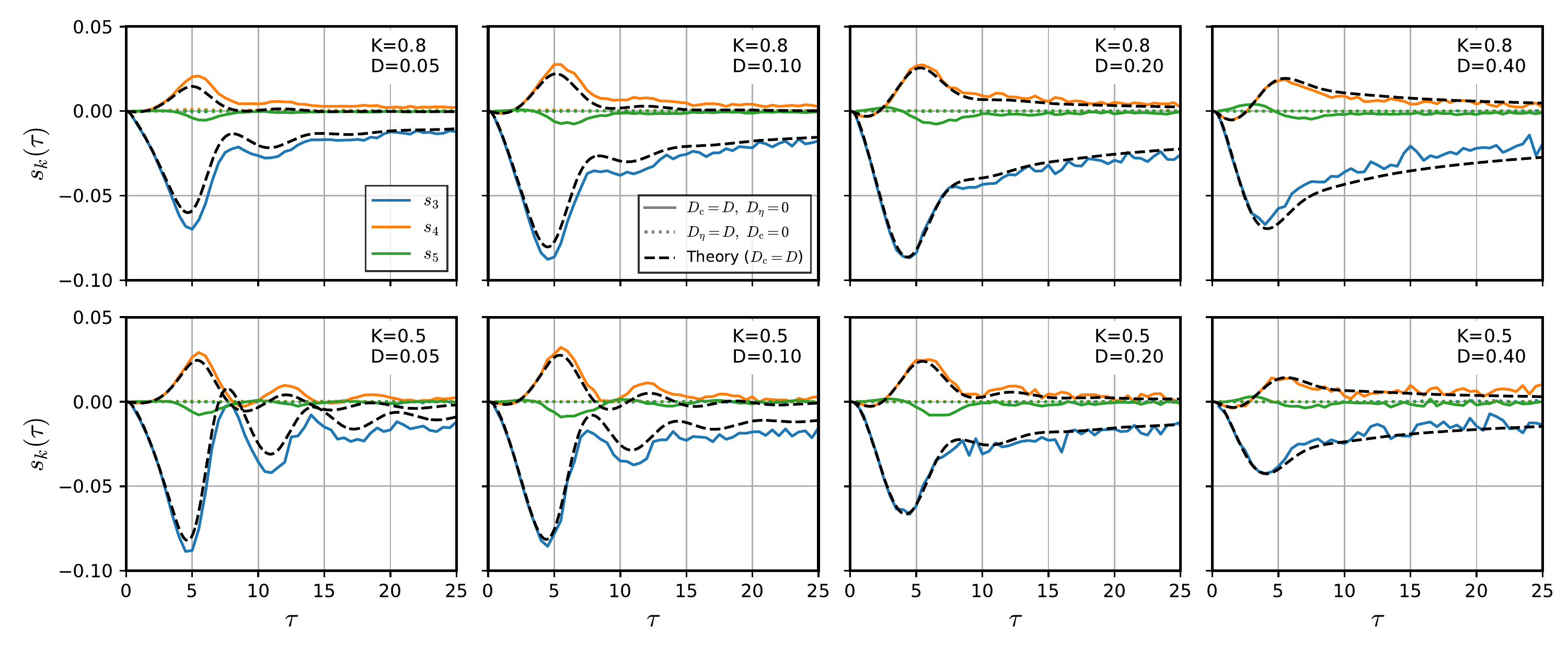}
\caption{Rescaled cumulants $s_3$ (blue), $s_4$ (orange) and $s_5$ as function of time $\tau$ for the case of common noise (solid lines) and purely individual noise (dotted lines) obtained from network simulations, shown together with the theoretical prediction for $s_3$ and $s_4$ the case of common noise (dashed black lines). Parameter values for $K$ and $D$ are indicated in the respective panels; other parameter values are $\omega_0=1$ and $\sigma_\omega=0$, and a simple sine coupling $f(\theta) = \sin\theta$ was used. The size of the simulated networks was $N=200$; further numerical parameters were as in Fig.~\ref{fig:ind_noise_Cx} except that $30$ networks were simulated for a duration of $10T_0$ each. 
\label{fig:cumulants}}
\end{figure*}

\subsubsection{Fourth cumulant}

In order to improve our description of the non-Gaussian statistics of $y_n(\tau)$, we also consider the fourth cumulant in the approximation of the average $\lr{e^{ily_n}}$, see Eq.~\eqref{eq:y_cumulants_approx}. The fourth cumulant of $y_n(\tau)$ is given by
\begin{align}
\kappa_4(\tau) &= \langle y_n^4(\tau;t)\rangle - 3\lr{y_n^2(\tau;t)}^2.
\end{align}
%where the second moment $\lr{y_n^2(\tau;t)}=2\Lambda(\tau) + 2(D_\eta + \Dc)\tau$ is already known in terms of the (as of yet undetermined) function $\Lambda$ and the noise intensities $D_\eta$ and $\Dc$. 
%We thus focus on the fourth moment
We first focus on the fourth moment
\begin{align}
\langle y_n^4(\tau;t)\rangle  &= \left\langle \prod_{j=1}^{4} \intl_{t}^{t+\tau} \!\!\! dt_j [\xi_{n}(t_{j})+\eta_{n}(t_{j})+\etac(t_{j})]\right\rangle .
\end{align}

Following the same reasoning as above for our calculation of the third cumulant, several terms below the quadruple integral can be discarded. First, all terms with odd powers of $\xi_n$ have to vanish because of the averaging properties of the connectivity matrix $K_{mn}$. In addition, all terms $\lr{\eta_n^p \etac^q}$, $p+q=4$, either vanish for odd powers $p$ and $q$ or are exactly compensated by the corresponding terms in the expression $3\lr{y_n^2(\tau)}^2$. (Note that $\eta_n$ and $\etac$ are Gaussian variables for which the fourth cumulant vanishes.) The remaining terms that potentially contribute to the fourth cumulant therefore are
%\begin{align}
%\kappa_4(\tau) &= \intl_t^{t+\tau}\!\!\!dt_1 \cdots \! \intl_t^{t+\tau}\!\!\!dt_4 \Bigg\{ \nn \\
%& \lr{\xi_n(t_1) \cdots \xi_n(t_4) } - 3\lr{\xi_n(t_1)\xi_n(t_2)}\lr{ \xi_n(t_3) \xi_n(t_4) } + \nn\\
%& 6 [ \lr{\xi_n(t_1)\xi_n(t_2) \eta_n(t_3) \eta_n(t_4)} - 3\lr{\xi_n(t_1)\xi_n(t_2)}\lr{ \eta_n(t_3) \eta_n(t_4) } + \nn\\
%& 6 [ \lr{\xi_n(t_1)\xi_n(t_2) \etac(t_3) \etac(t_4)} - 3\lr{\xi_n(t_1)\xi_n(t_2)}\lr{ \etac(t_3) \etac(t_4) } + \nn\\
%& 12 \lr{\xi_n(t_1)\xi_n(t_2) \eta_n(t_3) \etac(t_4)} 
%\end{align}
\begin{equation}
\label{eq:fourth_moment_terms}
\begin{aligned}
\lr{\xi_n(t_1) \cdots \xi_n(t_4) },\\ 
\lr{\xi_n(t_1)\xi_n(t_2) \eta_n(t_3) \eta_n(t_4)}, \\
 \lr{\xi_n(t_1)\xi_n(t_2) \etac(t_3) \etac(t_4)}, \\ 
 \lr{\xi_n(t_1)\xi_n(t_2) \eta_n(t_3) \etac(t_4)} .
\end{aligned}
\end{equation}
%$ \lr{\xi_n(t_1) \cdots \xi_n(t_4) }$, $\lr{\xi_n(t_1)\xi_n(t_2) \eta_n(t_3) \eta_n(t_4)} $, $\lr{\xi_n(t_1)\xi_n(t_2) \etac(t_3) \etac(t_4)}$, and $\lr{\xi_n(t_1)\xi_n(t_2) \eta_n(t_3) \etac(t_4)}$. 
We will briefly sketch the calculation for these terms but spare the reader the rather laborious calculations, and refer to appendix~\ref{app:fourth_cumul} for the details. 

First, we use again that the $\xi_n$ can be expressed in terms of the rotator phases, 
\begin{equation}
\xi_n(t)=\sum_{p,k} K_{np}|A_k|^2\exp(ik\theta_p(t)),
\end{equation}
and that $\theta_p(t) = \theta_p(t_0) + \int_{t_0}^tdt'(\omega_p+\xi_p+\eta_p+\etac)$. For all correlators containing the product $\xi_n(t_i)\xi_n(t_j)$, one eventually obtains expressions that involve the integrated input to the rotators $y_p(t_j-t_i)$, giving rise to averages of the type $\lr{e^{iky_p(t_2-t_1)}e^{ily_q(t_4-t_3)}}$,
%for the correlator $\lr{\xi_n(t_1) \cdots \xi_n(t_4) }$. Likewise,  for the correlators involving the external noise, one obtains expressions of the type 
$\lr{e^{iky_p(t_2-t_1)}\eta_n(t_3) \eta_n(t_4)}$, $\lr{e^{iky_p(t_2-t_1)}\etac(t_3) \etac(t_4)}$, and $\lr{e^{iky_p(t_2-t_1)}\eta_n(t_3) \etac(t_4)}$, respectively. For the last three contributions, we next use a similar identity as used above for the third cumulant (cf.~Eq.~\eqref{eq:exp_ab_rel1}):
\begin{align}
\label{eq:exp_abc_rel1}
\langle e^{ia} bc \rangle &= - \( \frac{d}{dr}\frac{d}{ds}  \left\langle e^{i(a+rb+sc)}\right\rangle \)_{r=0,s=0} .
\end{align}
In all four cases of Eq.~\eqref{eq:fourth_moment_terms}, we therefore  have to evaluate averages $\lr{e^{iz}}$, where for the first case $z=ky_p(t_2-t_1) + ly_q(t_4-t_3)$ for the first  and $z=a+rb+sc$ with $a=ky_p(t_2-t_1)$ and $b=\eta_{n/{\rm c}}(t_3)$, $c=\eta_{n/{\rm c}}(t_4)$ for the last three contributions. To compute these averages, we then make the second simplifying approximation discussed at the beginning of this section, effectively assuming Gaussian statistics for the stochastic variable $z$. In this approximation,  $\lr{e^{iz}}\approx e^{-\lr{z^2}/2}$, where only variances and covariances of $y_p$, $y_q$, $\eta_n$, and $\etac$ appear in $\lr{z^2}$, depending on the contribution we consider. For the last three terms, the Gaussian approximation directly leads to  
\begin{align}
\label{eq:exp_abc_rel2}
\langle e^{ia} bc \rangle &\approx \(\lr{bc} - \lr{ab}\lr{ac}\) e^{-\lr{a^2}/2} 
\end{align}
following relation \eqref{eq:exp_abc_rel1}, with appropriately chosen $a$, $b$, and $c$. 

Let us consider the first contribution, $\lr{\xi_n(t_1) \cdots \xi_n(t_4)}$, for which we have 
\begin{multline}
\lr{z^2}=k^{2}\lr{y_p^2(t_2-t_1)} +2kl\lr{y_p(t_2-t_1)y_q(t_4-t_3)}  \\
+ l^{2}\lr{y_q^2(t_4-t_3)}.
\end{multline}
For $p\neq q$, the covariance $\lr{y_p(t_2-t_1)y_q(t_4-t_3)}$ scales with $\Dc$ and vanishes in the absence of common input:
\begin{equation}
\lr{y_p(t_2-t_1)y_q(t_4-t_3)} = 2  \Dc \intl_{t_1}^{t_2}\!\!dt' \!\!\! \intl_{t_3}^{t_4}\!\!dt'' \delta(t''-t')
\end{equation}
For $p=q$, one can show that the covariances $\lr{y_p(t_2-t_1)y_q(t_4-t_3)}$ do not contribute to the final expression for $\lr{\xi_n(t_1) \cdots \xi_n(t_4)}$ as the average $\lr{e^{i(y_p+y_q)}}$ appears in a double sum over $p$ and $q$ and eventually scales with $1/N$, see appendix~\ref{app:fourth_cumul}.

Eventually, the first contribution can be written as
\begin{multline}
\lr{\xi_n(t_1) \cdots \xi_n(t_4)} \approx 3 K^4 \sum_{k,l}  g_k(t_2-t_1)g_l(t_4-t_3) \\
 \times \exp(- 2 kl \Dc \intl_{t_1}^{t_2}\!\!dt' \!\!\! \intl_{t_3}^{t_4}\!\!dt'' \delta(t''-t')) ,
\end{multline}
where we used the definition
\begin{equation}
\label{eq:def_gktau}
g_k(\tau) = |A_k|^2 \Phi(k\tau)e^{-k^2[\Lambda(\tau)+(D_\eta+\Dc)\tau]}
\end{equation}
to simplify the expression. For the fourth cumulant, we are eventually interested in the (integrated) difference
\begin{multline}
\kappa_{4,{\rm I}} = \intl_t^{t+\tau}\!\!\!dt_1 \cdots \! \intl_t^{t+\tau}\!\!\!dt_4 \big(\lr{\xi_n(t_1) \cdots \xi_n(t_4) }- \\
3\lr{\xi_n(t_1)\xi_n(t_2)}\lr{ \xi_n(t_3) \xi_n(t_4) }\big).
\end{multline}
After a somewhat lengthy calculation (see appendix~\ref{app:fourth_cumul}), we obtain an expression for the second derivative of $\kappa_{4,{\rm I}}$ that can be used for the computation of the fourth cumulant,
\begin{multline}
\frac{d^2}{d\tau^2}\kappa_{4,{\rm I}} =  24  K^4 \sum_{k,l}  \Bigg\{   \\
\intl_0^{\tau}dt    (\tau-t) g_k(\tau)  g_l(t) \left[ e^{-2kl\Dc t}  - 1\right] \ +  \\
\intl_{0}^\tau dt_a  \intl_{\tau-t_a}^{\tau}dt_b g_k(t_a)  g_l(t_b) \left[ e^{-2kl\Dc(t_a+t_b-\tau)} - 1 \right] \Bigg\} .
\end{multline}
Although not as conveniently integrated as the expression for the third cumulant, it still allows numerical solution. 

Based on relation~\eqref{eq:exp_abc_rel2}, we quickly discuss the three remaining contributions in Eq.~\eqref{eq:fourth_moment_terms}.
% $\lr{\xi_n(t_1)\xi_n(t_2) \eta_n(t_3) \eta_n(t_4)} $, $\lr{\xi_n(t_1)\xi_n(t_2) \etac(t_3) \etac(t_4)}$, and $\lr{\xi_n(t_1)\xi_n(t_2) \eta_n(t_3) \etac(t_4)}$. 
For $\lr{\xi_n(t_1)\xi_n(t_2) \eta_n(t_3) \eta_n(t_4)} $, since $\lr{y_p\eta_n}=0$ for $p\neq n$ (the covariances $\lr{ab}$ and $\lr{ac}$ of Eq.~\eqref{eq:exp_abc_rel2}), one can show that this term actually does not contribute to the fourth cumulant as it is exactly canceled by its `counterpart' in $3\mu_2^2$, see appendix~\ref{app:fourth_cumul}. The average $\lr{\xi_n(t_1)\xi_n(t_2) \eta_n(t_3) \etac(t_4)}$ does not have a corresponding term in $3\mu_2^2$, but vanishes nevertheless, as $\lr{y_p\eta_n}=0$ for $p\neq n$ (the covariance $\lr{ab}$ of Eq.~\eqref{eq:exp_abc_rel2}) and in addition $\lr{\eta_n\etac}=0$ (the covariance $\lr{bc}$ of Eq.~\eqref{eq:exp_abc_rel2}). Thus, the only remaining contribution to the fourth cumulant comes from $\lr{\xi_n(t_1)\xi_n(t_2) \etac(t_3) \etac(t_4)}$, for which we have to sum over terms of the type
\begin{multline}
\big( \lr{\etac(t_3)\etac(t_4)} - k^2 \lr{y_p(t_2-t_1)\etac(t_3)} \lr{y_p(t_2-t_1)\etac(t_4)} \big) \\
\times e^{-\frac{k^2\lr{y_p^2(t_2-t_1)}}{2}} ,
\end{multline}
see Eq.~\eqref{eq:exp_abc_rel2}. While the first term with the covariance $\lr{\etac(t_3)\etac(t_4)}$ is exactly cancelled by the corresponding product in $3\mu_2^2$, the second term does contribute to the fourth cumulant. After another somewhat lengthy calculation (see appendix~\ref{app:fourth_cumul}), the second derivative of this contribution to $\kappa_4$ is eventually found to be
\begin{align}
\frac{d^2}{d\tau^2}\kappa_{4,{\rm II}} &= - 48 \Dc^2  K^2 \sum_k  k^2  \tau^2 g_k(\tau).
\end{align}
As expected, this contribution vanishes in the absence of common noise ($\Dc=0$), consistent with deviations from Gaussian statistics being induced solely by input common to all units. The initial conditions for each of the two contributions to the fourth cumulant, $\kappa_{4,{\rm I}}(0)=0$ and $\kappa_{4,{\rm II}}(0)=0$ can be deduced from the defining quadruple integral;  the first time derivatives of $\kappa_{4,{\rm I}}$ and $\kappa_{4,{\rm II}}$ vanish as well for $\tau=0$, see appendix~\ref{app:fourth_cumul}. 

For the convenience of the reader, we provide below the final set of equations that define the self-consistent theory for the network noise autocorrelation function, including approximate corrections for non-Gaussian statistics up to the fourth cumulant:
\begin{widetext}
%\begin{align}
%\ddot\Lambda(\tau) &= K^2 \sum_l |A_l|^2  \Phi(l \tau) e^{-l^2[\Lambda(\tau) + (D_\eta + \Dc)\tau] -i\frac{il^3}{6}\kappa_3(\tau) + \frac{l^4}{24}\kappa_4(\tau)}  \\
%\ddot\kappa_3(\tau) &=12 \Dc K^2 \sum_l |A_l|^2 i l  \tau \Phi(l \tau) e^{-l^2[\Lambda(\tau) + (D_\eta + \Dc)\tau]} \\
%\ddot\kappa_4(\tau) &=24  K^4 \sum_{k,l}  \Bigg\{ \intl_0^{\tau} \!\! dt    (\tau-t) g_k(\tau)  g_l(t) \left[ e^{-2kl\Dc t}  - 1\right] \ + \intl_{0}^\tau \!\! dt_a \!\! \intl_{\tau-t_a}^{\tau}\!\!dt_b g_k(t_a)  g_l(t_b) \left[ e^{-2kl\Dc(t_a+t_b-\tau)} - 1 \right] \Bigg\}   \nn\\
%&\qquad\qquad - 48 \Dc^2  K^2 \sum_k  k^2  \tau^2 g_k(\tau) \\
%\Lambda(0)&=\dot\Lambda(0)=\kappa_3(0)=\dot\kappa_3(0)=\kappa_4(0)=\dot\kappa_4(0)=0.
%\end{align}
\begin{align}
\ddot\Lambda(\tau) &= K^2 \sum_l g_l(\tau) \exp\(-\frac{il^3}{6}\kappa_3(\tau) + \frac{l^4}{24}\kappa_4(\tau)\) \\
\ddot\kappa_3(\tau) &=12 \Dc K^2 \sum_l  i l  \tau g_l(\tau)  \\
\label{eq:table_kappa4_ddot}
\ddot\kappa_4(\tau) &=24  K^4 \sum_{k,l}  \Bigg\{ \intl_0^{\tau} \!\! dt    (\tau-t) g_k(\tau)  g_l(t) \left[ e^{-2kl\Dc t}  - 1\right] \ + \intl_{0}^\tau \!\! dt_a \!\! \intl_{\tau-t_a}^{\tau}\!\!dt_b g_k(t_a)  g_l(t_b) \left[ e^{-2kl\Dc(t_a+t_b-\tau)} - 1 \right] \Bigg\}   \nn\\
&\qquad - 48 \Dc^2  K^2 \sum_k  k^2  \tau^2 g_k(\tau) ,
\end{align} 
with the short-hand notation and initial conditions, respectively
\begin{align}
g_l(\tau) &= |A_l|^2 \Phi(l\tau) \exp\(-l^2[\Lambda(\tau)+(D_\eta+\Dc)\tau] \) ,\\
\Lambda(0)&=\dot\Lambda(0)=\kappa_3(0)=\dot\kappa_3(0)=\kappa_4(0)=\dot\kappa_4(0)=0 .
\end{align}
\end{widetext}
The network noise autocorrelation is then given as before by  $C_\xi(\tau)=\ddot\Lambda(\tau)$, see Eq.~\eqref{eq:Lambda}.

\subsection{Effects of common Gaussian white noise}

We now turn to the effects that are specific to the common noise and are captured with our modified theory. For a fixed total noise intensity $D_\eta+\Dc=D=\text{const}$, we thus compare the cases of purely intrinsic noise ($D_\eta=D$, $\Dc=0$) and purely common noise ($\Dc=D$, $D_\eta=0$). We will restrict ourselves to a pure sine coupling $f(\cdot)=\sin(\cdot)$ and vary the coupling strength $K$ and total noise intensity $D$. 

\subsubsection{Common Gaussian white noise induces non-Gaussian network fluctuations}

We have already seen in our calculation that the network fluctuations cannot be any longer assumed to be Gaussian in the presence of input shared among the different units. In Fig.~\ref{fig:cumulants} we confirm this theoretical finding by numerical simulations, and also illustrate that our approximations for the third and fourth cumulants work reasonably well for the considered parameters. The cumulants are functions of the time argument $\tau$, and are shown in Fig.~\ref{fig:cumulants} in a rescaled version that capture the deviation from Gaussianity relative to the (time-dependent) standard deviation $\sigma(\tau)=\kappa_2^{1/2}(\tau)$ of the distribution of the $y_n$. This corresponds to rewriting the expansion Eq.~\eqref{eq:y_cumulants} as 
\begin{equation*}
\langle e^{i l y_n(\tau)} \rangle = \exp\(\sum_{k=1}^{\infty}(il\sigma(\tau))^{k}s_{k}\),
\end{equation*}
where $s_k = \kappa_k/(\kappa_2^{k/2}k!)$ are the rescaled cumulants shown in Fig.~\ref{fig:cumulants}. Because $\lr{y_n}=0$ and by the definition of the variance, we have $s_1\equiv 0$ and $s_2\equiv 1/2$. For $k>2$, the $s_k$ can be thought of as generalizations of the skewness $\kappa_3/\kappa_2^{3/2}$ and excess kurtosis $\kappa_4/\kappa_2^2$ to higher orders, weighted by  $1/k!$ with which the $k$th cumulant enters the series Eq.~\eqref{eq:y_cumulants}. For common noise,  $s_3(\tau)$, $s_4(\tau)$ and $s_5(\tau)$ (solid lines) are different from zero, illustrating that the deviation from Gaussianity is significant. However, they remain always small compared to $s_2\equiv1/2$, and especially $s_5$ is small. This lends further support to the basic idea of our approximation, i.e., the expansion around the Gaussian case and the neglect of even higher cumulants. 

For purely intrinsic noise, the magnitudes of the rescaled higher-order cumulants (dotted lines in Fig.~\ref{fig:cumulants}) are extremely small -- expected fluctuations due to finite-size effects cannot even be resolved on the scale of this figure and are zero for all practical purposes. This is a striking confirmation of our assumption of Gaussian network noise statistics for the case of purely individual noise.

\begin{figure}
\includegraphics[width=0.8\columnwidth]{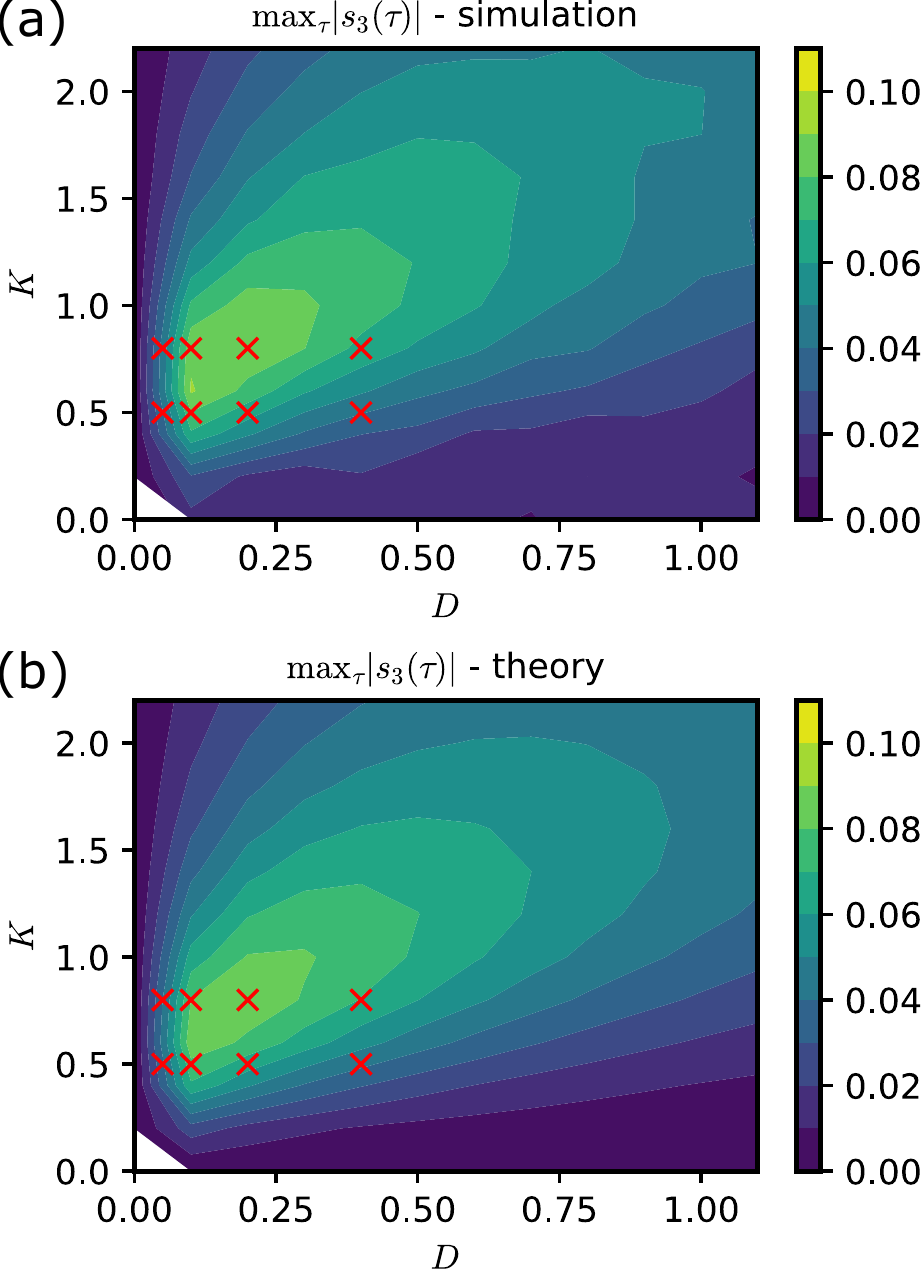}
\caption{The maximum rescaled skewness $\max{}_\tau |s_3(\tau)|$ obtained from network simulations (a) and from our theory (b) as a function of coupling strength $K$ and common noise strength $D$. Other parameters as in Fig.~\ref{fig:cumulants}. Crosses indicate the parameter values used in the different panels of Fig.~\ref{fig:cumulants}.
\label{fig:KD_s3}}
\end{figure}

\begin{figure}
\includegraphics[width=0.8\columnwidth]{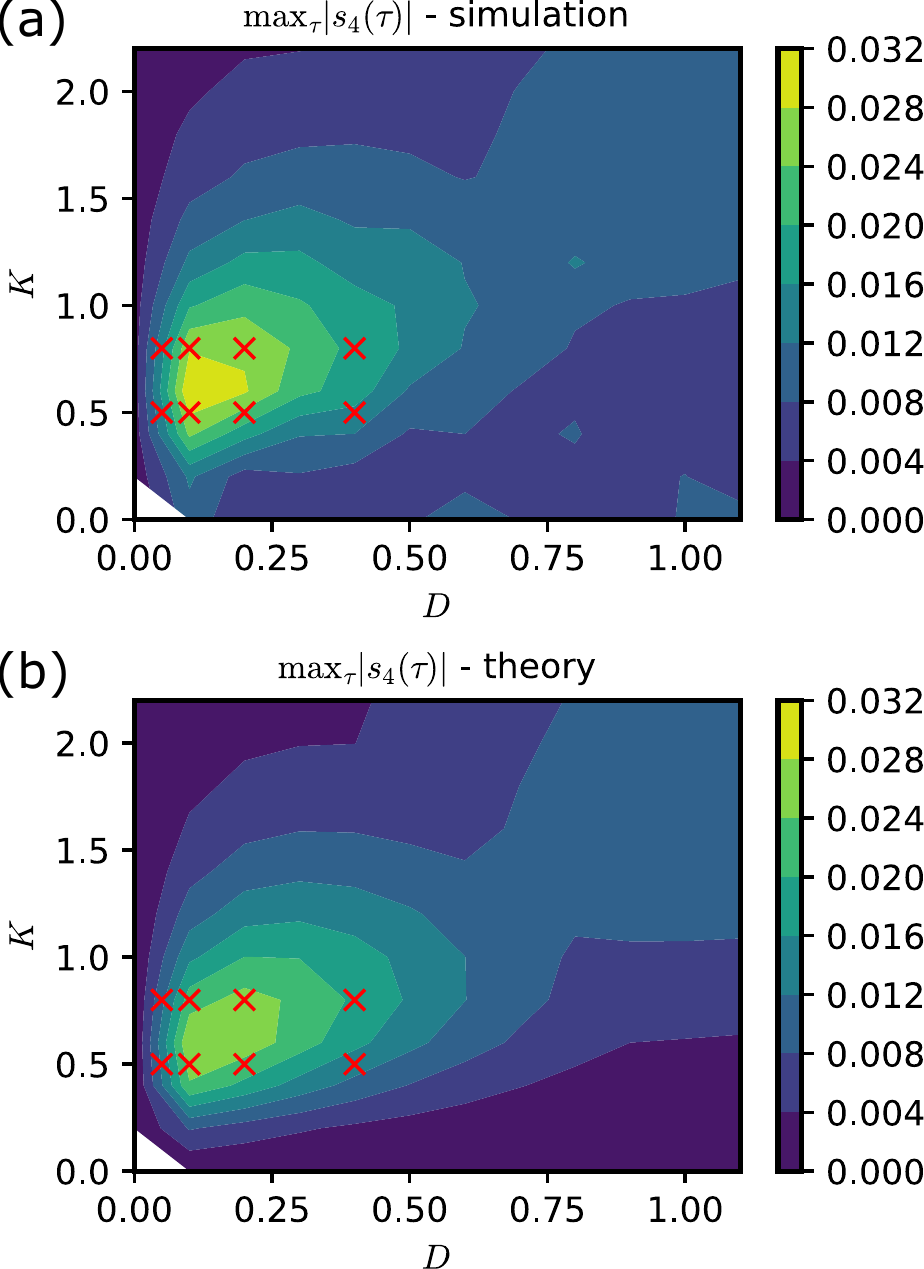}
\caption{The maximum rescaled excess kurtosis $\max{}_\tau |s_4(\tau)|$ obtained from network simulations (a) and from our theory (b) as a function of coupling strength $K$ and common noise strength $D$. Other parameters as in Fig.~\ref{fig:cumulants}. Crosses indicate the parameter values used in the different panels of Fig.~\ref{fig:cumulants}.
\label{fig:KD_s4}}
\end{figure}

Our theory that involves a number of approximations provides a quantitatively reasonable description of the numerically measured cumulants for a variety of parameters, irrespective of the intensity of the common noise $D$ and the strength of the coupling $K$, Fig.~\ref{fig:cumulants}. In particular, the theory reproduces the pronounced oscillations at low total noise (e.g.~$K=0.5$ and $D=0.05$ of Fig.~\ref{fig:cumulants}e) and the respective maxima of the time-dependent functions at all parameters considered. Small deviations are observed at intermediate times. 

To provide a more comprehensive overview of the dependence on the parameters, we plot the maxima of $|s_3(\tau)|$ and $|s_4(\tau)|$ as functions of $K$ and $D$, in Figs.~\ref{fig:KD_s3} and~\ref{fig:KD_s4}, respectively. The quality of our approximation becomes apparent from the comparison of the simulation results (panels a) and the theoretical predictions (panels b). Remarkably, these functions display global maxima at non-vanishing but finite values of $K$ and $D$, indicating a maximal deviation from Gaussianity at this parameter combination ($K_{\rm max} \approx 0.6$ and $D_{\rm max} \approx 0.1$ within our resolution for both $|s_3(\tau)|$ and $|s_4(\tau)|$ and both simulations and theory). 
%In the limit of our resolution, the maxima of $|s_3(\tau)|$ and $|s_4(\tau)|$ in both simulations and theory are attained at $K_\max\approx\0.6$ and $D_\max\approx0.1$. 
Specifically, $s_3$ and $s_4$ seem to vanish for all limiting cases of vanishing or infinite $K$ or $D$, which is consistent with the the integrated input $y_n$ becoming effectively Gaussian. In the following, we argue why this is so.  

For $D\to 0$, $y_n$ becomes Gaussian because here the system approaches the previously studied purely autonomous network, in which fluctuations are Gaussian~\cite{vanMeegen:2018hj}. For $K\to0$, the integrated inputs $y_n$ is completely dominated by the common noise that is Gaussian by assumption. Thus, it is also plausible that the cumulants vanish in this limit. For $D\to\infty$ at fixed $K$, a similar argument holds true: The common noise will dominate the integrated input, in which the network noise is limited in amplitude, hence also in this limit the fluctuations will be Gaussian. Finally, for $K\to\infty$ at fixed $D$, the dynamics of the network approaches that of a purely autonomous network without external input, simply because the network fluctuations are orders of magnitude stronger than the common noise.

In summary, significantly non-vanishing values of $s_3$ and $s_4$ appear as predicted by the theory when the network is subject to a common noise, whereas a purely individual noise cannot evoke any non-Gaussian features of the network fluctuations. Our theory captures the non-Gaussian statistics reasonably well.

\begin{figure*}
\includegraphics[width=0.8\textwidth]{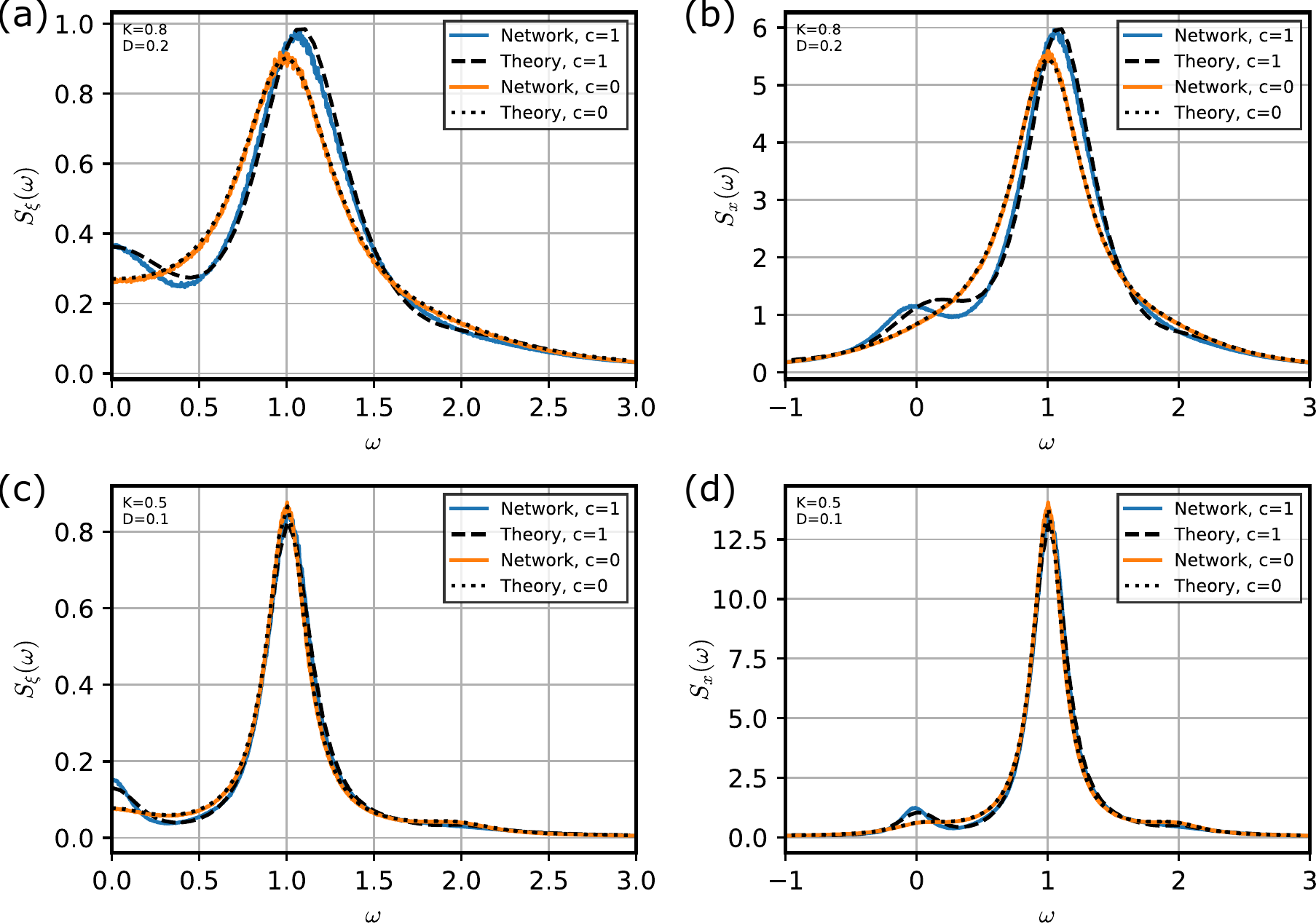}
\caption{Power spectra $S_\xi$ and $S_x$ of the network noise $\xi_m$ (a,c) and the rotators $x_m$ (b,d), respectively, for purely common ($c=1$, i.e.~$\Dc=D,D_\eta=0$) and purely intrinsic ($c=0$ i.e.~$\Dc=0,D_\eta=D$) noise. (a,b) $K=0.8$, $D=0.2$, (c,d) $K=0.5$, $D=0.1$. Network simulations (solid lines) are compared with the theoretical prediction (dashed and dotted lines). The power spectra for the theory were computed from the numerical solution until $t_{\rm num}=125$. Other parameters as in Fig.~\ref{fig:cumulants}.
\label{fig:spectra}}
\end{figure*}

\begin{figure}
\includegraphics[width=0.8\columnwidth]{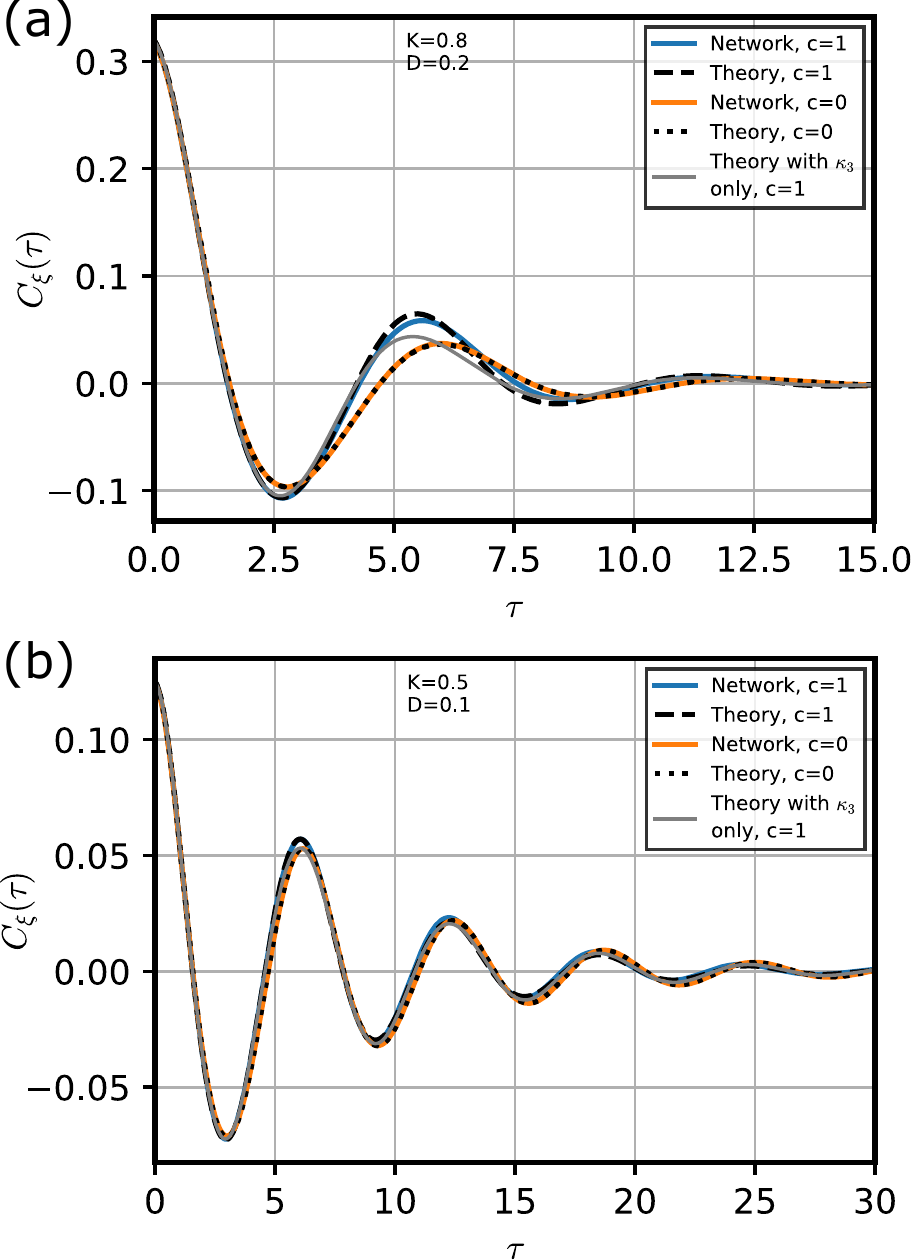}
\caption{
\label{fig:correlations_common}
Correlation function $C_\xi(\tau)$ of the network noise $\xi_m$ for strong (a) and weak (b) noise. Line style, color code, and parameters as in Fig.~\ref{fig:spectra}.  
The solution of the simpler system obtained when including only the third cumulant, Eqs.~\eqref{eq:ddot_kappa3} and~\eqref{eq:ddot_Lambda_3cumul_only} is additionally shown (thin grey line) for comparison.}
\end{figure}

\subsubsection{Common Gaussian white noise increases spectral low-frequency power and shifts peaks to higher frequencies}

We can use the computed cumulants to obtain the correlation functions and, by an additional Fourier transform, the power spectra of the network noise $\xi_i$ and the individual pointers $x_i$. We look at those statistics for two selected points in the parameter plane ($K$,$D$), respectively (0.8, 0.2) and (0.5,0.1), see Fig.~\ref{fig:spectra}. Two effects of the common noise can be observed if we compare to the case of purely individual noise of the same strength. Firstly, there is an increase of power around zero frequency, both for the network fluctuations $\xi_m(t)$ (left) and the individual rotators $x_m$ (right). This increase is roughly a factor of two for $K=0.5$, $D=0.1$, but smaller for the other set. Because the external noise has power for all frequencies, this increase in low frequency power is a non-trivial prediction of our theory that is confirmed in the simulations. Power at low frequencies corresponds to slow fluctuations, which thus seem to become more important in the presence of common noise. The described effect stands in marked contrast to that observed in populations of spiking neurons with global inhibition, in which common noise can induce a spectral peak at non-vanishing frequencies and a reduction of power around zero frequency~\cite{Doiron:2004cj,Lindner:2005hd}.

%Discussion slow mode for coupled maps? Our results are closer to that obtained for coupled 
%Discussion slow fluctuations more generally? ... that can emerge also by other mechanisms \cite{Ostojic2014,Wieland2015}

There is a second effect of the common noise on the spectra. For $K=0.8$, $D=0.2$, we observe that the main spectral peak is shifted from the intrinsic rotator frequency $\omega_0$ to a somewhat higher frequency; the same effect is in principle also present at the other parameter set but much weaker. The increase in frequency corresponds to a slightly faster oscillation of the units. It  is also predicted by our theory and confirmed by the simulations.

For completeness, in Fig.~\ref{fig:correlations_common} we show for the two parameter sets the self-consistent autocorrelation functions $C_{\xi}(\tau)$ of the network fluctuations and $C_{x}(\tau)$ of the rotators, which allow for another comparison between theory and simulations. Indeed, for both parameter sets, deviations of the simulations from the theory appear to be minor, and taking into account only the first four cumulants yields a good agreement. 
%In particular for $K=0.8$, $D=0.2$, deviations in the spectral domain seem to be more pronounced. 
We also take the opportunity to show how the theory would work if we would only include the third cumulant. Deviations from the simulations are stronger when only $\kappa_3$ is used, and there is a clear benefit to include the fourth cumulant.

%They can be directly obtained via Eqs.~\eqref{} and~\eqref{} once $\Lambda$, $\kappa_3$ and $\kappa_4$ have been calculated.  

\section{Summary \& outlook}
\label{sec:discussion}

In this paper, we have studied a network of randomly coupled phase oscillators which are driven by intrinsic and external (common) noise. We have generalized the theory of self-consistent correlation functions developed in Ref.~\onlinecite{vanMeegen:2018hj} for the case of the autonomous network (i.e.~in the absence of stochastic forcing). It turned out that this generalization is rather straightforward in the case of purely intrinsic (private) noise. Here, the network fluctuations are still to a very good approximation Gaussian -- in fact, they approach Gaussian statistics in the strict thermodynamic limit of an infinite network. Our theory works very well in describing the autocorrelation functions of the network noise and of the individual rotators. Furthermore, we also compared the situations in which all rotators receive a private noise of a given intensity and in which only the observed rotator is subject to noise of that intensity. The network-mediated effect of the private noise on all rotators may have nontrivial consequences on the correlation statistics of the observed rotator. The coherence of the oscillation may decrease or increase by stochastic forcing of the rest of the network, depending on the chosen parameters and the coupling function. 

When the rotators are subject to common noise, the picture becomes more complicated. We have shown that, somewhat paradoxically, the driving with a common Gaussian noise turns a Gaussian network noise into a non-Gaussian one. This is reminiscent of earlier observations made in the case of coupled chaotic maps, where the mean-field ceases to be Gaussian in the thermodynamic limit~\cite{Kaneko:1990gx,Kaneko:1992}; an effect which is reproduced when independent maps are subject to a common noise~\cite{Griniasty:1994ht}. For the rotator network, the non-Gaussian statistics demand a serious revision of the theory developed by van Meegen \& Lindner~\cite{vanMeegen:2018hj}. Here we put forward a perturbative approach based on a cumulant expansion around the Gaussian case that leads to a system of few equations for the network noise autocorrelation and the third and fourth (time-dependent) cumulants of the integrated network noise. The latter measure the deviations from Gaussianity and vanish in the absence of common noise. We have evaluated these equations for large ranges of coupling strength $K$ and noise intensity $\Dc$ for a simple sine coupling. We have found that our theory describes the network statistics in most cases surprisingly well. Notably, looking at the temporal maximum of the absolute values of third and fourth cumulants, we found non-monotonic dependences of these statistics on $K$ and $\Dc$. Specifically, there are non-vanishing values $K^*$ and $\Dc^*$ for which the rescaled skewness $s_3$ and  excess kurtosis $s_4$ are maximal. Nonetheless, even in this case, the non-Gaussian deviations are moderate and still captured by our theory. We recall that the latter is based on two important assumptions: i) We consider higher-order cumulants only up to the fourth cumulant, and ii) assume Gaussian statistics in the calculations of the (non-Gaussian) third and fourth cumulants of the integrated network noise. In principle, it is possible to relax these assumptions and to calculate the fifth cumulant and/or consider non-Gaussian contributions in the calculation of the cumulants. However, the calculations would become even more involved and, at least for our system, the agreement between predicted and measured correlation functions (the actual statistics of interest) is already satisfactory with the approximations made.
We note in passing that the expansion in cumulants we propose here is somehow remindful of the approach presented in Refs.~\onlinecite{Tyulkina:2018kb,Goldobin:2018}, where circular cumulants were used to improve the prediction of the order parameter for Kuramoto oscillators with intrinsic noise and Lorentz-distributed frequencies. These works were not concerned with the statistics of the fluctuating ``recurrent'' input however, and the first Kuramoto order parameter being finite, only the second cumulant was considered as a correction in applications. 
%\rot{Die zweite Kumulante ist die Varianz - das waere also keine Korrektur zur Gauss-Statistik sondern innerhalb der Gauss-Statistik. Ist das wirklich das, was Du sagen willst?}

%However, the comparison of measured and predicted cumulants indicate that the latter assumption suffices to describe their temporal dynamics. 

%The comparison of the power spectra and correlation functions of network noise and ... approximation fourth cumulant. 

What are the effects of the non-Gaussian network fluctuations induced by common noise at the level of the correlation functions and power spectra? We found two effects on the spectra. For once, the common noise may shift the main peak to higher frequencies. Secondly, the common Gaussian white, i.e. temporally uncorrelated, noise gives rise to increased low-frequency power in both network noise and individual rotator spectra. Both of these effects are not easily explained, but are well captured by our self-consistent theory of non-Gaussian fluctuations. It is still an open problem to extract these features from our set of equations by analytical techniques, e.g.~by an appropriately simplified Fourier transformation of the equations in simple limit cases as was done in~\cite{vanMeegen:2018hj}. We note that slow fluctuations have also been observed to arise spontaneously in recurrent networks of LIF neurons with sufficiently strong (current-based) synaptic coupling in the absence of common noise~\cite{Ostojic:2014kd,Wieland:2015fx} and one may speculate how common input may further increase low frequency power in those networks. On a different note, the increase in low-frequency power in our system stands in marked contrast to the reduction of low-frequency power due to common noise in populations of integrate-and-fire neurons with global inhibitory feedback~\cite{Doiron:2004cj,Lindner:2005hd}.  
%Interestingly, deviations  

Our theory may serve as a template for calculations of the self-consistent correlation statistics for other systems of interest. A straightforward generalization of our results is possible, for instance if the common noise is temporally correlated, as characterized by a correlation function $\Cc(\tau)$. More complicated will be the case of a coupling matrix that respects Dale's law, i.e.~that a given unit $j$ is either excitatory ($K_{ij}>0$ for all $i$) or inhibitory ($K_{ij}<0$), possibly combined with a given connection probability $p$ between units. Preliminary simulation results show that in this case, the statistics of network fluctuations cease to be Gaussian. Given the importance of this constraint in the neural context, a generalization of our theory to this case is a particularly promising topic of future research.

%\begin{itemize}
%
%
%\item Intrinsic noise of networks still hot topic
%\item Gaussian statistics often supposed (non-Gaussianity has been observed but to my knowledge not been pursued further)
%\item LÃ©vy-noise resulting from microscopic dynamics/forces (Chandrasekar?)
%\item ``Violation of the law of large numbers'' --- Kaneko, PRL 65, p.~1391, 1990 \& subsequent studies (emergent dynamisc in coupled maps etc., see also later work on Kuramoto models that cite these papers)
%\item Recently proposed model of rotator networks promising framework to gain analytic insight!
%\end{itemize}

\appendix
\section{Detailed calculation of the fourth cumulant}
\label{app:fourth_cumul}

Here we aim to provide some details of the calculation of the fourth cumulant $\kappa_4(\tau)$ of the stochastic variable $y_n(\tau)$. Since $\lr{y_n(\tau)}=0$, the fourth cumulant is given by $\kappa_4 = \lr{y_n^4} - 3\lr{y_n^2}^2$. For the fourth moment, we have
\begin{widetext}
\begin{align}
\langle y_n^4(\tau)\rangle &= \left\langle \prod_{j=1}^{4} \intl_{t}^{t+\tau} \!\!\! dt_j [\xi_{n}(t_{j})+\eta_{n}(t_{j})+\etac(t_{j})]\right\rangle \\
&= \intl_t^{t+\tau}\!\!\!dt_1 \cdots \! \intl_t^{t+\tau}\!\!\!dt_4 \big(\lr{\prod_{j=1}^{4}\xi_n(t_j)} + \lr{\prod_{j=1}^{4}\eta_n(t_j)} + \lr{\prod_{j=1}^{4}\etac(t_j)} + 12 \lr{\xi_n(t_1)\xi_n(t_2)\eta_n(t_3) \etac(t_4)} \nn \\
&\qquad + 6 \lr{\xi_n(t_1)\xi_n(t_2)\eta_n(t_3) \eta_n(t_4)}  + 6 \lr{\xi_n(t_1)\xi_n(t_2)\etac(t_3) \etac(t_4)} + 6 \lr{\eta_n(t_1) \eta_n(t_2)\etac(t_3) \etac(t_4)}  \big),
\end{align}
\end{widetext}
where we already used that odd powers of $\xi_n$ vanish when averaging over the Gaussian, zero-centered connectivity matrix $K_{mn}$. (We also used that $\lr{\eta_n^3\etac}=\lr{\eta_n^3}\lr{\etac}=0$, $\lr{\eta_n\etac^3}=\lr{\eta_n}\lr{\etac^3}=0$ as both noise sources are uncorrelated by definition.) Of note, we can easily match all terms but one to a corresponding term in $3\lr{y_n^2}^2$:
\begin{widetext}
\begin{align}
3\langle y_n^2(\tau)\rangle^2 &= 3 \intl_t^{t+\tau}\!\!\!dt_1 \cdots \! \intl_t^{t+\tau}\!\!\!dt_4 \left\langle \prod_{j=1}^{2}  [\xi_{n}(t_{j})+\eta_{n}(t_{j})+\etac(t_{j})]\right\rangle  \left\langle \prod_{j=3}^{4}  [\xi_{n}(t_{j})+\eta_{n}(t_{j})+\etac(t_{j})]\right\rangle  \\
%&= \intl_t^{t+\tau}\!\!\!dt_1 \cdots \! \intl_t^{t+\tau}\!\!\!dt_4 \big( 3 \lr{\xi_n(t_1)\xi_n(t_2)} \lr{\xi_n(t_3)\xi_n(t_4)} +  3 \lr{\eta_n(t_1)\eta_n(t_2)} \lr{\eta_n(t_3)\eta_n(t_4)} + 3 \lr{\etac(t_1)\etac(t_2)} \lr{\etac(t_3)\etac(t_4)} \nn \\  
&= \intl_t^{t+\tau}\!\!\!dt_1 \cdots \! \intl_t^{t+\tau}\!\!\!dt_4 \big( 3 \lr{\prod_{j=1}^{2}\xi_n(t_j)} \lr{\prod_{j=3}^{4}\xi_n(t_j)} +  3 \lr{\prod_{j=1}^{2}\eta_n(t_j)} \lr{\prod_{j=3}^{4}\eta_n(t_j)} + 3 \lr{\prod_{j=1}^{2}\etac(t_j)} \lr{\prod_{j=3}^{4}\etac(t_j)} \nn \\  
&\qquad + 6 \lr{\xi_n(t_1)\xi_n(t_2)}\lr{\eta_n(t_3) \eta_n(t_4)}  + 6 \lr{\xi_n(t_1)\xi_n(t_2)}\lr{\etac(t_3) \etac(t_4)} + 6 \lr{\eta_n(t_1) \eta_n(t_2)}\lr{\etac(t_3) \etac(t_4)}  \big).
\end{align}
\end{widetext}
For a process $y_n(\tau)$ with perfectly Gaussian statistics, this is exactly the expression the fourth moment would boil down to, and $\mu_4 - 3\mu_2^2=0$. Here, we are looking for a difference arising in the fourth moment; this is slightly different from the third cumulant which---for a centered process---is directly given by the third moment. All terms that only involve powers of $\eta_n$ and $\etac$ are  purely Gaussian by definition, and the differences of the corresponding terms vanish. We are left with four potentially finite contributions,
\begin{widetext}
\begin{multline}
\langle y_n^4(\tau)\rangle - 3\langle y_n^2(\tau)\rangle^2  = \intl_t^{t+\tau}\!\!\!dt_1 \cdots \! \intl_t^{t+\tau}\!\!\!dt_4 \Big[
\underbrace{\lr{\prod_{j=1}^{4}\xi_n(t_j)} -  3 \lr{\prod_{j=1}^{2}\xi_n(t_j)} \lr{\prod_{j=3}^{4}\xi_n(t_j)} }_{\rm I}
 + 12 \underbrace{\lr{\xi_n(t_1)\xi_n(t_2)\eta_n(t_3) \etac(t_4)}}_{\rm II}  \\
+ 6 \big(\underbrace{\lr{\xi_n(t_1)\xi_n(t_2)\eta_n(t_3) \eta_n(t_4)}  - \lr{\xi_n(t_1)\xi_n(t_2)}\lr{\eta_n(t_3) \eta_n(t_4)}}_{\rm III}  \big)
+ 6 \big(\underbrace{\lr{\xi_n(t_1)\xi_n(t_2)\etac(t_3) \etac(t_4)} -\lr{\xi_n(t_1)\xi_n(t_2)}\lr{\etac(t_3) \etac(t_4)} }_{\rm IV}  \big) \Big],
\end{multline}
\end{widetext}
which we discuss separately in the following. We just mention that we can in the following assume $t_2>t_1$ for all terms without loss of generality, and consider the fourfold integral over the domain $\intl_t^{t+\tau}\!\!\!dt_1  \intl_{t_1}^{t+\tau}\!\!\!dt_2 \intl_t^{t+\tau}\!\!\!dt_3  \intl_{t}^{t+\tau}\!\!\!dt_4$ with the integrands being multiplied by two.  For term I, we can and will additionally assume $t_4>t_3$ without loss of generality for the same symmetry reasons, and accordingly the multiply the integrand by four.

\subsection{$\lr{\prod_{j=1}^{4}\xi_n(t_j)} -  3 \lr{\prod_{j=1}^{2}\xi_n(t_j)} \lr{\prod_{j=3}^{4}\xi_n(t_j)}$}

By expressing the network noise again in terms of the couplings $K_{mn}$ and phases $\theta_n$ based on the Fourier decomposition of the coupling function,
\begin{equation}
\xi_n(t) = \sum_m K_{nm} \sum_l A_l e^{ik\(\theta_p(t_0) + \int_{t_0}^t dt' \dot\theta_m(t')\)} , 
\end{equation}
we obtain the following expression for the first term:
\begin{widetext}
\begin{align}
\langle \xi_n(t_1)\cdots \xi_n(t_4)\rangle\  &= \left\langle\sum_{m,o,p.q} K_{nm}K_{no}K_{np}K_{nq}
\sum_{k,l,r,s}A_kA_lA_rA_s e^{ik\[\theta_m(t_0)+\int_{t_0}^{t_1}dt'\dot\theta_m(t')\]}\cdots e^{is\[\theta_q(t_0)+\int_{t_0}^{t_4}dt'\dot\theta_q(t')\]} \right\rangle \\
&= \sum_{m,o,p,q} \sum_{k,l,r,s} A_kA_lA_rA_s \left\langle K_{nm}K_{no}K_{np}K_{nq} \right\rangle_K \left\langle e^{ik\[\theta_m(t_0)+\int_{t_0}^{t_1}dt'\dot\theta_m(t')\]}\cdots e^{is\[\theta_q(t_0)+\int_{t_0}^{t_4}dt'\dot\theta_q(t')\]} \right\rangle .
\end{align}
\end{widetext}

We separately average over the disorder, which in the case of Gaussian couplings gives 
\begin{align}
\left\langle K_{nm}K_{no}K_{np}K_{nq} \right\rangle_K &= \frac{K^4}{N^2}\(\delta_{mo}\delta_{pq}+\delta_{mp}\delta_{oq}+\delta_{mq}\delta_{op}\).
\end{align}
After eliminating the $\delta_{\cdot\cdot}\delta_{\cdot\cdot}$ by summing over two of the four corresponding indices $m$, $o$, $p$ and $q$ and appropriate renaming of the remaining indices, one thus obtains
\begin{widetext}
\begin{align}
&\langle \xi_n(t_1)\cdots \xi_n(t_4)\rangle  = \frac{K^4}{N^2} \sum_{p,q} \sum_{k,l,r,s} A_kA_lA_rA_s \Bigg[\nn\\
&\Bigg\langle e^{ik\[\theta_p(t_0)+\int_{t_0}^{t_1}dt'\dot\theta_p(t')\]} e^{il\[\theta_p(t_0)+\int_{t_0}^{t_2}dt'\dot\theta_p(t')\]}  e^{ir\[\theta_q(t_0)+\int_{t_0}^{t_3}dt'\dot\theta_q(t')\]} e^{is\[\theta_q(t_0)+\int_{t_0}^{t_4}dt'\dot\theta_q(t')\]} \Bigg\rangle+\nn\\
&\Bigg\langle e^{ik\[\theta_p(t_0)+\int_{t_0}^{t_1}dt'\dot\theta_p(t')\]} e^{il\[\theta_p(t_0)+\int_{t_0}^{t_3}dt'\dot\theta_p(t')\]}  e^{ir\[\theta_q(t_0)+\int_{t_0}^{t_2}dt'\dot\theta_q(t')\]} e^{is\[\theta_q(t_0)+\int_{t_0}^{t_4}dt'\dot\theta_q(t')\]} \Bigg\rangle+\nn\\
& \Bigg\langle e^{ik\[\theta_p(t_0)+\int_{t_0}^{t_1}dt'\dot\theta_p(t')\]} e^{il\[\theta_p(t_0)+\int_{t_0}^{t_4}dt'\dot\theta_p(t')\]}  e^{ir\[\theta_q(t_0)+\int_{t_0}^{t_2}dt'\dot\theta_q(t')\]} e^{is\[\theta_q(t_0)+\int_{t_0}^{t_3}dt'\dot\theta_q(t')\]} \Bigg\rangle \Bigg] .
\end{align}
\end{widetext}
The three terms differ in their combinations of the time arguments $t_1$, $t_2$, $t_3$ and $t_4$, and in principle cannot be subsumed into a single term. As they appear below a quadruple integral over all four time arguments, we can for the ease of notation and without loss of generality permute time indices separately for the three terms however, and consider 
\begin{widetext}
\begin{align}
\langle \xi_n(t_1)\cdots\xi_n(t_4)\rangle &= 3 \frac{K^4}{N^2}  \sum_{p,\ldots,s} A_kA_lA_rA_s 
\Bigg\langle e^{ik\[\theta_p(t_0)+\int_{t_0}^{t_1}dt'\dot\theta_p(t')\]} e^{il\[\theta_p(t_0)+\int_{t_0}^{t_2}dt'\dot\theta_p(t')\]} 
 \cdots e^{is\[\theta_q(t_0)+\int_{t_0}^{t_4}dt'\dot\theta_q(t')\]} \Bigg\rangle \nn\\
&=3 \frac{K^4}{N^2}  \sum_{p,\ldots,s} A_kA_lA_rA_s 
\Bigg\langle e^{i(k+l)\theta_p(t_0)} e^{ik\int_{t_0}^{t_1}dt'\dot\theta_p(t')} e^{il\int_{t_0}^{t_2}dt'\dot\theta_p(t')} 
 e^{i(r+s)\theta_q(t_0)} e^{ir \int_{t_0}^{t_3}dt'\dot\theta_q(t')} e^{is\int_{t_0}^{t_4}dt'\dot\theta_q(t')} \Bigg\rangle .
\end{align}
\end{widetext}

We next average over random initial phases $\theta_p(t_0)$, $\theta_p(t_0)$ at the reference time $t_0$, with   
$\left\langle e^{i(k+l)\theta_p(t_0)}\right\rangle_{\theta_p(t_0)}  = \delta_{k,-l}$ and the analogous relation for $\theta_q(t_0)$,
%$$\left\langle e^{i(r+s)\theta_q(t_0)}\right\rangle_{\theta_q(t_0)}  &= \delta_{r,-s},$$
leading to 
\begin{widetext}
\begin{align}
\langle \xi_m(t_1)\cdots \xi_m(t_4)\rangle &= 3  \frac{K^4}{N^2}  \sum_{p,q}\sum_{k,l} |A_k|^2 |A_l|^2  \left\langle e^{ik\int_{t_1}^{t_2}dt'\dot\theta_p(t')} e^{il\int_{t_3}^{t_4}dt'\dot\theta_q(t')} \right\rangle .
\end{align}
\end{widetext}
Using $\dot \theta_m = \omega_m + \xi_m + \eta_m + \etac$ and averaging separately over the intrinsic frequencies, $\lr{e^{ix\omega_i}}_{\bs{\omega}}=\Phi(x)$ being the characteristic function of the $\omega_i$, one eventually obtains
\begin{widetext}
\begin{align}
\langle \xi_m(t_1)\cdots \xi_m(t_4)\rangle&= 3  \frac{K^4}{N^2}  \sum_{p,q}\sum_{k,l} |A_k|^2 |A_l|^2  \left\langle e^{ik\omega_p(t_2-t_1)}\right\rangle  \left\langle e^{il\omega_q(t_4-t_3)}\right\rangle \times \nn\\
&\qquad\qquad\qquad\qquad \left\langle e^{ik\int_{t_1}^{t_2}dt'[\xi_p(t') + \eta_p(t') + \etac(t')]} e^{il\int_{t_3}^{t_4}dt'[\xi_q(t') + \eta_q(t') + \etac(t')]} \right\rangle \\
\label{eq:app_k4_I_sm}
&= 3 \frac{K^4}{N^2}  \sum_{p,q}\sum_{k,l} |A_k|^2 |A_l|^2  \Phi(k(t_2-t_1))  \Phi(l(t_4-t_3))  \lr{e^{i [k y_p(t_2-t_1;t_1) + ly_q(t_4-t_3;t_3)]}} ,
\end{align}
\end{widetext}
where we furthermore used the definition of Eq.~\eqref{eq:y_def_common} for the $y_p$, $y_q$ that appear in the average in the last equation. 

In principle, the average $\lr{e^{i [k y_p(t_2-t_1;t_1) + ly_q(t_4-t_3;t_3)]}}$ can again be expressed in terms of the cumulants of $z=k y_p(t_2-t_1;t_1) + ly_q(t_4-t_3;t_3)$, in perfect analogy to Eq.~\eqref{eq:y_cumulants}. However, we will make here again the second approximation of our theory and consider only the second cumulant of $z$ when evaluating the average $\lr{e^{iz}}$, effectively assuming that the deviations from Gaussianity of the $y_p$ and $y_q$ are small and can be neglected in the calculation of the higher-order corrections to the average $\lr{e^{iy_n}}$, i.e.~the third and fourth cumulant of $y_n$. This approximation allows one to express $\lr{e^{i [k y_p(t_2-t_1;t_1) + ly_q(t_4-t_3;t_3)]}}$ only in terms of the variances and covariances of $y_p$ and $y_q$,
\begin{multline}
\lr{e^{i z}} = e^{-\frac{l^2}{2}\lr{y_p^2(t_2-t_1;t_1)}} e^{- \frac{k^2}{2}\lr{y_q^2(t_4-t_3;t_3)}}\times\\ 
e^{- kl\lr{y_p(t_2-t_1;t_1)y_q(t_4-t_3;t_3)} },
\end{multline}
which we can calculate within the self-consistent theory for the correlation function of the network noise. In particular, the covariance is given by
\begin{widetext}
\begin{align}
\lr{y_p(t_2-t_1;t_1)y_q(t_4-t_3;t_3)} &=  \intl_{t_1}^{t_2}\!\!dt' \intl_{t_3}^{t_4}\!\!dt''\lr{[\xi_p(t') + \eta_p(t') + \etac(t')][\xi_q(t'') + \eta_q(t'') + \etac(t'')] }\\
\label{eq:app_k4_I_ypyq}
&=  \intl_{t_1}^{t_2}\!\!dt' \intl_{t_3}^{t_4}\!\!dt'' \[ \delta_{pq} \(C_\xi(t''-t') + C_\eta(t''-t')\) + \Cc(t''-t')\] .
\end{align}
\end{widetext}
(Note that  the crosscorrelation $\lr{\xi_p(t')\xi_q(t'')}$ vanishes for $p\neq q$, as can be seen from $\sum_{m,n}\lr{K_{pm}K_{qn} f(\theta_m(t'))f(\theta_n(t''))} = \sum_{m,n}\lr{K_{pm}K_{qn}}\lr{ f(\theta_m(t'))f(\theta_n(t''))}$ and $\lr{K_{pm}K_{qn}}=0$ when $p\neq q$.)

The double sum over $p$ and $q$ of Eq.~\eqref{eq:app_k4_I_sm} contains $N$ terms with $p=q$ and $N(N-1)$ terms with $p\neq q$, while the sum has a prefactor $1/N^2$. In the large-$N$ limit, we neglect all contributions that scale with $1/N$ and keep thus only the last term in Eq.~\eqref{eq:app_k4_I_ypyq}. With $\Cc(t''-t') = 2\Dc\delta(t''-t')$ and $\lr{y_{\{p,q\}}^2(\tau,t)} = 2\Lambda(\tau) + 2(D_\eta + \Dc)\tau$ (see Eq.~\eqref{eq:kappa2_common}), we thus obtain 
\begin{multline}
\langle \xi_n(t_1)\cdots \xi_n(t_4)\rangle  = 3  K^4 \sum_{k,l}  g_k(t_2-t_1)  g_l(t_4-t_3)  \times\\
 e^{-2kl\Dc \intl_{t_1}^{t_2}dt' \intl_{t_3}^{t_4}dt'' \delta(t''-t') }  ,
\end{multline}
where we used the definition for the $g_i(t)$ given by Eq.~\eqref{eq:def_gktau}.

Along the same lines, it is straightforward to show that the corresponding term from $\lr{y_n^2(\tau)}^2$  is given by
\begin{equation}
\lr{\xi_n(t_1)\xi_n(t_2)} \lr{\xi_n(t_3) \xi_n(t_4)} = K^4 \sum_{k,l}  g_k(t_2-t_1)  g_l(t_4-t_3) ,
\end{equation}
so that we eventually obtain the following expression for the first contribution to the fourth cumulant:
\begin{widetext}
\begin{align}
\kappa_{4,{\rm I}}(\tau)  &= \intl_t^{t+\tau}\!\!\!dt_1 \cdots \! \intl_t^{t+\tau}\!\!\!dt_4 \(\lr{\prod_{j=1}^{4}\xi_n(t_j)} -  3 \lr{\prod_{j=1}^{2}\xi_n(t_j)} \lr{\prod_{j=3}^{4}\xi_n(t_j)} \) \nn \\
&= 12 K^4 \sum_{k,l} \intl_t^{t+\tau}\!\!\!dt_1  \intl_{t_1}^{t+\tau}\!\!\!dt_2 \intl_t^{t+\tau}\!\!\!dt_3  \intl_{t_3}^{t+\tau}\!\!\!dt_4 
g_k(t_2-t_1)  g_l(t_4-t_3) \(  e^{-2kl\Dc \intl_{t_1}^{t_2}dt' \intl_{t_3}^{t_4}dt'' \delta(t''-t') } - 1 \) .
\end{align}
\end{widetext}
With the variable transformations $t_a=t_2-t_1$, $t_b = t_4-t_3$, and $t_c=t_3-t_1$ we can rewrite this expression as 
\begin{widetext}
\begin{align}
\kappa_{4,{\rm I}}(\tau)  &= 12 K^4 \sum_{k,l} \intl_0^\tau\!\!dt_a  \intl_0^\tau\!\!dt_b \! \intl_{t_a-\tau}^{\tau-t_b}\!\!\!dt_c \,  g_k(ta)  g_l(t_b) \(  e^{-2kl\Dc \intl_0^{t_a}dt' \intl_{t_c}^{t_c + t_b}dt''\delta(t''-t') } - 1 \) \intl_{t_{\rm low}(\tau,t_a,t_b,t_c)}^{t_{\rm upp}(\tau,t_a,t_b,t_c)}\!\!\!dt_1 ,
\end{align}
\end{widetext}
where care has to be taken in identifying the correct integration limits $t_{\rm low}(\tau,t_a,t_b,t_c)$ and $t_{\rm upp}(\tau,t_a,t_b,t_c)$ for $t_1$ when changing the order of integration. Making first a case distinction for $t_a>t_b$ and $t_b>t_a$, respectively, and then considering $t_c$ relative to $t_a-t_b$ and 0, we eventually find 
\begin{widetext}
\begin{align}
\label{eq:app_k4_4_t1_cases}
\intl_{t_{\rm low}(\tau,t_a,t_b,t_c)}^{t_{\rm upp}(\tau,t_a,t_b,t_c)}dt_1 
&= \begin{cases}
\tau-t_a+t_c & t_a - \tau < t_c < \min(t_a-t_b,0) \\
\tau-\max(t_a,t_b) & \min(t_a-t_b,0)  < t_c < \max(t_a-t_b,0)  \\
\tau-t_b-t_c & \max(t_a-t_b,0)  <t_c<\tau-t_b 
\end{cases} .
\end{align}
\end{widetext}

While the numerical evaluation of the resulting triple integral is possible \emph{a priori}, we try and simplify the expression further by taking the second time derivative with respect to $\tau$. For the first time derivatives w.r.t.~the integration limits of the triple integral over $t_a$, $t_b$, and $t_c$, one can see that they vanish when we use the respective values for  $t_a$, $t_b$, and $t_c$ with Eq.~\eqref{eq:app_k4_4_t1_cases}. The remaining time derivative of $\int_{t_{\rm low}}^{t_{\rm upp}}dt_1$ is surprisingly simple and equal to 1, as can be seen as well from Eq.~\eqref{eq:app_k4_4_t1_cases}. Thus, 
\begin{widetext}
\begin{align}
\frac{d}{d\tau}\kappa_{4,{\rm I}}(\tau)  &= 12 K^4 \sum_{k,l} \intl_0^\tau\!\!dt_a  \intl_0^\tau\!\!dt_b \! \intl_{t_a-\tau}^{\tau-t_b}\!\!\!dt_c \,  g_k(t_a)  g_l(t_b) \(  e^{-2kl\Dc G(t_a,t_b,t_c) } - 1 \), 
\end{align}
\end{widetext}
where we furthermore introduced the short-hand notation
\begin{equation}
G(t_a,t_b,t_c) = \int_0^{t_a}dt' \int_{t_c}^{t_c + t_b}dt''\delta(t''-t') .
\end{equation}
The second time derivative is then given by 
\begin{widetext}
\begin{align}
\frac{d^2}{d\tau^2}\kappa_{4,{\rm I}}(\tau)  
%&= 12 K^4 \sum_{k,l} \frac{d}{d\tau} \intl_0^\tau\!\!dt_a  \intl_0^\tau\!\!dt_b \! \intl_{t_a-\tau}^{\tau-t_b}\!\!\!dt_c \,  g_k(t_a)  g_l(t_b) \(  e^{-2kl\Dc G(t_a,t_b,t_c) } - 1 \), \\
&= 12 K^4 \sum_{k,l} \Big[ \int_0^{\tau}dt_b    \int_{0}^{\tau-t_b}dt_c  g_k(\tau)  g_l(t_b) \( e^{-2kl\Dc G(\tau,t_b,t_c) }  - 1\) + \nn\\
&\qquad\qquad \int_0^{\tau}dt_a  \int_{t_a-\tau}^0dt_c  g_k(t_a)  g_l(\tau) \( e^{-2kl\Dc G(t_a,\tau,t_c) }  - 1\)  + \nn\\
&\qquad\qquad \int_0^{\tau}dt_a \int_0^{\tau}dt_b    g_k(t_a)  g_l(t_b) \( e^{-2kl\Dc G(t_a,t_b,\tau-t_b) }  - 1\)  + \nn\\
&\qquad\qquad \int_0^{\tau}dt_a \int_0^{\tau}dt_b    g_k(t_a)  g_l(t_b) \( e^{-2kl\Dc G(t_a,t_b,t_a-\tau) }  - 1\) \Big] .
\end{align}
\end{widetext}
Let us consider the four contributions separately. 

\begin{widetext}
1) $t_a=\tau:  \quad t_c>0,\  t_c+t_b<\tau \ \Rightarrow \ G(\tau,t_b,t_c) =  t_b$
\begin{equation}
\int_0^{\tau}dt_b \int_{0}^{\tau-t_b}dt_c  g_k(\tau)  g_l(t_b) \( e^{-2kl\Dc G(\tau,t_b,t_c) }  - 1\)   
= \int_0^{\tau}dt_b    (\tau-t_b) g_k(\tau)  g_l(t_b) \( e^{-2kl\Dc t_b}  - 1\) 
\end{equation} 

2) $t_b=\tau:  \quad  t_c<0,\  t_c+\tau>t_a \  \Rightarrow \  G(t_a,\tau,t_c) =  t_a$

\begin{equation}
\int_0^{\tau}dt_a  \int_{t_a-\tau}^0dt_c  g_k(t_a)  g_l(\tau) \( e^{-2kl\Dc G(t_a,\tau,t_c) }  - 1\) =  \int_0^{\tau}dt_a  (\tau-t_a) g_k(t_a)  g_l(\tau) \( e^{-2kl\Dc t_a }  - 1\) 
\end{equation} 
%\end{widetext}
Note that this term is completely analogous to the previous one and because of the $k,l$ symmetry both can be combined by simply computing one of them and taking it twice.

%\begin{widetext}
3) $t_c = \tau-t_b:  \quad  t_c>0,\  t_c+t_b>t_a \ \Rightarrow \  G(t_a,t_b,\tau-t_b) = \begin{cases}
0 & t_a+t_b<\tau \ (t_c>t_a) \\
t_a+t_b-\tau & t_a+t_b>\tau
\end{cases}$
\begin{align}
\int_0^{\tau}dt_a \int_0^{\tau}dt_b    g_k(t_a)  g_l(t_b) \( e^{-2kl\Dc G(t_a,t_b,\tau-t_b) }  - 1\) = \int_{0}^\tau dt_a  \int_{\tau-t_a}^{\tau}dt_b g_k(t_a)  g_l(t_b) \( e^{-2kl\Dc (t_a+t_b-\tau)} - 1 \)
\end{align}

4) $t_c = t_a-\tau:  \quad t_c<0,\ t_c+t_b<t_a \ \Rightarrow \ G(t_a,t_b,t_a-\tau) = \begin{cases}
0 & t_a+t_b<\tau \ (t_c+t_b<0) \\
t_a+t_b-\tau & t_a+t_b>\tau
\end{cases}$

\begin{align}
\int_0^{\tau}dt_a \int_0^{\tau}dt_b    g_k(t_a)  g_l(t_b) \( e^{-2kl\Dc G(t_a,t_b,t_a-\tau) }  - 1\) = \int_{0}^\tau dt_a  \int_{\tau-t_a}^{\tau}dt_b g_k(t_a)  g_l(t_b) \( e^{-2kl\Dc (t_a+t_b-\tau)} - 1 \)
\end{align}
This term is obviously identical to the previous one.
\end{widetext}

Based on the above, the four terms can be regrouped into a simple integral from $0\ldots\tau$ and a double integral,
\begin{widetext}
\begin{align}
\frac{d^2}{d\tau^2}\kappa_{4,{\rm I}}(\tau)   &=  24  K^4 \sum_{k,l}  \[   \intl_0^{\tau}dt_b    (\tau-t_b) g_k(\tau)  g_l(t_b) \( e^{-2kl\Dc t_b}  - 1\) + \intl_{0}^\tau dt_a  \intl_{\tau-t_a}^{\tau}\!dt_b g_k(t_a)  g_l(t_b) \( e^{-2kl\Dc (t_a+t_b-\tau)} - 1 \) \] ,
\end{align}
\end{widetext}
which is the first contribution to the expression given in Eq.~\eqref{eq:table_kappa4_ddot}.

\subsection{$\lr{\xi_n(t_1)\xi_n(t_2)\eta_n(t_3) \etac(t_4)}$}

We can express the  $\xi_n(t_j)$ in terms of the coupling matrix $K_np$ and the phases $\theta_p$, and arrive along the very same lines as above at the following expression:
%\begin{widetext}
\begin{multline}
\label{eq:app_k4_II_sum_pk}
\lr{\xi_n(t_1)\xi_n(t_2)\eta_n(t_3) \etac(t_4)} = \\ 
\frac{K^2}{N}  \sum_p\sum_k |A_k|^2 \Phi(k(t_2-t_1)) \times\\
 \lr{e^{i k y_p(t_2-t_1;t_1)}  \eta_n(t_3) \etac(t_4) } .
\end{multline}
%\end{widetext}

To evaluate the remaining average $\lr{e^{i k y_p(t_2-t_1;t_1)}  \eta_n(t_3) \etac(t_4) }$, we first use relation~\eqref{eq:exp_abc_rel1} to express it as a derivative w.r.t.~$r$ and $s$ of an average of the type $\lr{e^{iz}}$, where $z=y_p(t_2-t_1;t_1) + r\eta_n(t_3) + s \etac(t_4)$.  Here, we make again our second approximation and neglect higher-order cumulants of $z$ when evaluating $\lr{e^{iz}}$, effectively treating $z$ as Gaussian for the calculation of the higher-order cumulants of $y_n(\tau)$. We can than express $\lr{e^{i k y_p(t_2-t_1;t_1)}  \eta_n(t_3) \etac(t_4) }$ in terms of the variances and covariances of $y_p(t_2-t_1;t_1)$, $\eta_n(t_3)$, and $\etac(t_4)$, see Eq.~\eqref{eq:exp_abc_rel2}, and obtain
\begin{multline}
\lr{e^{i k y_p(t_2-t_1;t_1)}  \eta_n(t_3) \etac(t_4) } =
\big( \lr{\eta_n(t_3)\etac(t_4)} - \\
k^2 \lr{y_p(t_2-t_1)\eta_n(t_3)} \lr{y_p(t_2-t_1)\etac(t_4)} \big) \\
\times e^{-\frac{k^2\lr{y_p^2(t_2-t_1)}}{2}} .
\end{multline}
However, upon closer inspection one finds that none of these terms contributes in the large-$N$ limit, as $\lr{\eta_n(t_3)\etac(t_4)} =0$ by definition and
\begin{equation}
\lr{y_p(t_2-t_1)\eta_n(t_3)} = \delta_{pn} 2D_\eta \Theta(t_3-t_1) \Theta(t_2-t_3).
\end{equation}
The sum over $p$ in Eq.~\eqref{eq:app_k4_II_sum_pk} thus reduces to a single term that scales with $1/N$,
\begin{widetext}
\begin{align}
\label{eq:app_k4_II_final}
\lr{\xi_n(t_1)\xi_n(t_2)\eta_n(t_3) \etac(t_4)} = 
4 D_\eta \Dc \frac{K^2}{N}  \sum_k g_k(t_2-t_1) k^2 \Theta(t_3-t_1) \Theta(t_2-t_3) 
\Theta(t_4-t_1) \Theta(t_2-t_4) , 
\end{align}
\end{widetext}
and vanishes as $N\to\infty$.

\subsection{$\lr{\xi_n(t_1)\xi_n(t_2)\eta_n(t_3) \eta_n(t_4)}  - \lr{\xi_n(t_1)\xi_n(t_2)}\lr{\eta_n(t_3) \eta_n(t_4)}$}

We can directly apply the intermediate results from the previous contribution, in particular relation~\eqref{eq:exp_abc_rel2} also discussed in the main text, and write
\begin{widetext}
\begin{align}
\label{eq:app_k4_III_1}
\lr{\xi_n(t_1)\xi_n(t_2)\eta_n(t_3) \eta_n(t_4)} &= \frac{K^2}{N}  \sum_p \sum_k |A_k|^2 \Phi(k(t_2-t_1))  \lr{e^{i k y_p(t_2-t_1;t_1)}  \eta_n(t_3) \eta_n(t_4) } \\
\label{eq:app_k4_III_sum}
&=\frac{K^2}{N}  \sum_p \sum_k g_k(t_2-t_1) \big( \lr{\eta_n(t_3)\etac(t_4)} - k^2 \lr{y_p(t_2-t_1;t_1)\eta_n(t_3)} \lr{y_p(t_2-t_1;t_1)\eta_n(t_4)} \big) .
\end{align}
\end{widetext}
The first term is exactly canceled by $ \lr{\xi_n(t_1)\xi_n(t_2)}\lr{\eta_n(t_3) \eta_n(t_4)}$ which we need to subtract to compute the cumulant. The averages $lr{y_p(t_2-t_1;t_1)\eta_n(t_3)}$ contribute only for $p=n$ as discussed above, and we are left with
\begin{widetext}
\begin{multline}
\label{eq:app_k4_III_2}
\lr{\xi_n(t_1)\xi_n(t_2)\eta_n(t_3) \eta_n(t_4)} - \lr{\xi_n(t_1)\xi_n(t_2)}\lr{\eta_n(t_3) \eta_n(t_4)} \\
= - 4 D_\eta^2 \frac{K^2}{N}  \sum_k k^2 g_k(t_2-t_1) k^2 \Theta(t_3-t_1) \Theta(t_2-t_3) \Theta(t_4-t_1) \Theta(t_2-t_4) .
\end{multline}
\end{widetext}
This term also vanishes in the limit of $N\to\infty$ and does not contribute to the fourth cumulant.

\subsection{$\lr{\xi_n(t_1)\xi_n(t_2)\eta_c(t_3) \eta_c(t_4)}  - \lr{\xi_n(t_1)\xi_n(t_2)}\lr{\eta_c(t_3) \eta_c(t_4)}$}

This last term follows again straightforwardly from the previous contribution, where we have simply to replace $\eta_n$ by $\etac$. However, the averages $\lr{y_p(t_2-t_1;t_1)\eta_n(t_j)}$ contribute for each $p$ in the sum which is the equivalent of Eq.~\eqref{eq:app_k4_III_sum}, and we eventually obtain 
\begin{widetext}
\begin{multline}
\label{eq:app_k4_IV_1}
\lr{\xi_n(t_1)\xi_n(t_2)\etac(t_3) \etac(t_4)} - \lr{\xi_n(t_1)\xi_n(t_2)}\lr{\etac(t_3) \etac(t_4)} \\
= - 4 \Dc^2 K^2  \sum_k k^2 g_k(t_2-t_1) \Theta(t_3-t_1) \Theta(t_2-t_3) \Theta(t_4-t_1) \Theta(t_2-t_4) .
\end{multline}
\end{widetext}
The contribution to the fourth cumulant is accordingly given by
\begin{widetext}
\begin{align}
\kappa_{4,{\rm IV}}(\tau)  &= 6 \intl_t^{t+\tau}\!\!\!dt_1 \cdots \! \intl_t^{t+\tau}\!\!\!dt_4 \(\lr{\xi_n(t_1)\xi_n(t_2)\eta_c(t_3) \eta_c(t_4)}  - \lr{\xi_n(t_1)\xi_n(t_2)}\lr{\eta_c(t_3) \eta_c(t_4)} \) \nn \\
&= - 48 \Dc^2 K^2 \sum_k k^2 \intl_t^{t+\tau}\!\!\!dt_1  \intl_{t_1}^{t+\tau}\!\!\!dt_2  g_k(t_2-t_1) (t_2-t_1)^2 .
\end{align}
\end{widetext}

To further simplify the calculation of $\kappa_{4,{\rm IV}}(\tau)$, we again take the second time derivative w.r.t.~$\tau$. After the variable substitution $t_a=t_2-t_1$ and changing the integration to $ \int_t^{t+\tau}dt_1  \int_{0}^{t+\tau-t_1}dt_a$, it is straightforward to obtain 
\begin{align}
\frac{d}{d\tau} \kappa_{4,{\rm IV}}(\tau)  &= - 48 \Dc^2 K^2 \sum_k k^2  \intl_0^\tau\!dt' g_k(t') t'^2 ,
\end{align}
where we furthermore used the substitution $t'=t+\tau-t_1$. The second time derivative then follows immediately as  
\begin{align}
\frac{d^2}{d\tau^2} \kappa_{4,{\rm IV}}(\tau)  &= - 48 \Dc^2 K^2 \sum_k k^2  g_k(\tau) \tau^2 ,
\end{align}
which is the second contribution to the expression given in Eq.~\eqref{eq:table_kappa4_ddot}.

%\section{Numerical methods}

%\bibliography{aipsamp}% Produces the bibliography via BibTeX.

%\bibliographystyle{apsrev}
%\nocite{*}
%\bibliography{ALL_22_03_19,rotator_refs}
\bibliography{rotator_refs}

\end{document}